\documentclass[sigconf]{acmart}

\usepackage{tabularray}
\usepackage{tabularx} 
\usepackage{graphicx} 
\usepackage{booktabs}
\usepackage{wrapfig}
\usepackage{lipsum}
\usepackage{caption} 
\usepackage{enumitem}
\usepackage{soul}

\definecolor{highlightcolor}{HTML}{CCFF99}
\sethlcolor{highlightcolor}

\AtBeginDocument{%
  \providecommand\BibTeX{{%
    \normalfont B\kern-0.5em{\scshape i\kern-0.25em b}\kern-0.8em\TeX}}}

\copyrightyear{2024}
\acmYear{2024}
\setcopyright{rightsretained}
\acmConference[UIST '24]{The 37th Annual ACM Symposium on User Interface Software and Technology}{October 13--16, 2024}{Pittsburgh, PA, USA}
\acmBooktitle{The 37th Annual ACM Symposium on User Interface Software and Technology (UIST '24), October 13--16, 2024, Pittsburgh, PA, USA}
\acmDOI{10.1145/3654777.3676368}
\acmISBN{979-8-4007-0628-8/24/10}

\begin{document}


\title{{\tool:} Grounding Natural Language Database Queries with Interactive Explanations}

%

\author{Yuan Tian}
\affiliation{%
  \institution{Purdue University}
  \city{West Lafayette}
  \state{IN}
  \country{USA}
}
\email{tian211@purdue.edu}

\author{Jonathan K. Kummerfeld}
\affiliation{%
  \institution{University of Sydney}
  \city{Sydney}
  \country{Australia}
}
\email{jonathan.kummerfeld@sydney.edu.au}

\author{Toby Jia-Jun Li}
\affiliation{%
  \institution{University of Notre Dame}
  \city{Notre Dame}
  \state{IN}
  \country{USA}
}
\email{toby.j.li@nd.edu}

\author{Tianyi Zhang}
\affiliation{%
  \institution{Purdue University}
  \city{West Lafayette}
  \state{IN}
  \country{USA}
}
\email{tianyi@purdue.edu}

%


\newcommand{\edit}[1]{#1}

\newcommand{\tool}{\textsc{SQLucid}}
\newcommand{\participants}{30}
\newcommand{\oldtool}{\textsc{Steps}}
\newcommand{\redtodo}[1]{\textcolor{red}{#1}}
\newcommand{\todo}[1]{\textcolor{red}{#1}}
\newcommand{\bluetodo}[1]{\textcolor{blue}{#1}}
\newcommand{\edited}[1]{\textcolor{black}{#1}}
\newcommand{\tlcomment}[1]{\noindent{\\\textcolor{magenta}{\textbf{\#\#\# TL:}
\textsf{#1} \#\#\#\\}}}

 \newcommand{\circled}[1]{{\large \textcircled{\footnotesize #1}}}

\newcommand{\scorecolor}[1]{%
    \ifnum#1=0 \cellcolor{green!0}#1%
    \else\ifnum#1<4 \cellcolor{green!20}#1%
    \else\ifnum#1<8 \cellcolor{green!40}#1%
    \else\ifnum#1<12 \cellcolor{green!60}#1%
    \else \cellcolor{green!80}#1%
    \fi\fi\fi\fi%
}


\begin{abstract}
Though recent advances in machine learning have led to significant improvements in natural language interfaces for databases, the accuracy and reliability of these systems remain limited, especially in high-stakes domains.
This paper introduces {\tool}, a novel user interface that bridges the gap between non-expert users and complex database querying processes. 
{\tool} addresses existing limitations by integrating visual correspondence, intermediate query results, and editable step-by-step \edit{SQL explanations in natural language} to facilitate user understanding and engagement. 
This unique blend of features empowers users to understand and refine SQL queries easily and precisely. 
Two user studies and one quantitative experiment were conducted to validate {\tool}'s effectiveness, showing significant improvement in task completion accuracy and user confidence compared to existing interfaces. 
\edit{Our code is available at} \url{https://github.com/magic-YuanTian/SQLucid}.

\end{abstract}



\begin{teaserfigure}
    \centering
    \includegraphics[width=0.9\linewidth]{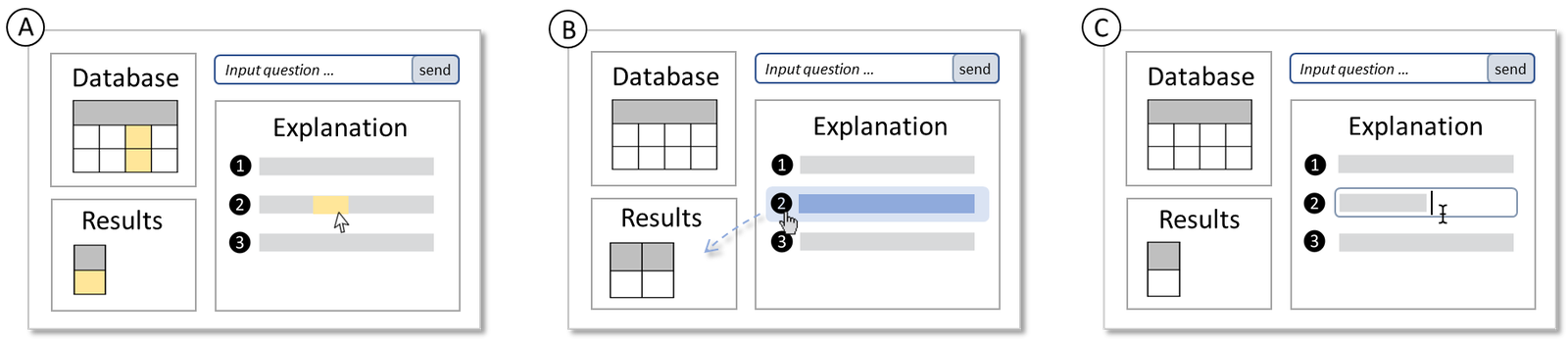}
    \caption{An Overview of {\tool}: 
        \textcircled{\scriptsize A} Given a model-generated SQL query, {\tool} helps the user understand the query behavior by generating a step-by-step explanation in natural language. When the user hovers over an entity (e.g., a column name) in the natural language explanation, {\tool} will highlight the corresponding elements in the database and query result to help the user grasp the visual correspondence between the explanation and the database. 
        \textcircled{\scriptsize B} For each step in the natural language explanation, the user can inspect the intermediate query result of the step to validate query behavior and diagnose query errors.
        \textcircled{\scriptsize C} Once the user identifies the erroneous step, they can directly edit the explanation of that step to specify the correct behavior and guide the model to refine the query.
    } 
    \Description{...}
    \label{fig:teaser}
\end{teaserfigure}

\received{20 February 2007}
\received[revised]{12 March 2009}
\received[accepted]{5 June 2009}

\maketitle

\section{Introduction}

The rise of big data has led to a growing demand for querying databases for data analysis and decision-making. 
To fully unleash the analytical power of databases, many natural language (NL) interfaces~\cite{lunar, diy, splash, dialsql} have been developed, enabling non-experts to express and fulfill their goals through NL queries.
The backbone of these interfaces is a computational approach that translates an NL query to a database query in a formal language such as SQL. Early work in this domain applied rule-based or grammar-based approaches~\cite{zelle1996learning, kate2005learning}.
Recent advances in deep learning have led to a variety of text-to-SQL models~\cite{editsql, ratsql, sqlnet, dinsql}, achieving unprecedented performance on NL querying tasks. 

Despite these great strides, text-to-SQL models cannot always reliably generate correct queries aligned with user intent.
As a result, users run the risk of receiving wrong query results and henceforth making incorrect or suboptimal decisions. This is critical in high-stakes domains such as finance and healthcare. 
The leaderboard of a popular evaluation text-to-SQL benchmark, Spider\footnote{\url{https://yale-lily.github.io/spider}\label{leaderboard}}, indicates that
even with the best system~\cite{dinsql} built on GPT-4 still suffers from an error rate of 10\%. 
It is crucial to help users identify and fix the potential errors in the database queries generated by these models to avoid incorrect or suboptimal decisions. \edit{To bridge the gap, several approaches have been developed to enable users to provide feedback to SQL generation in an interactive manner}~\cite{binary, construct_interface, misp, diy, piia}.

\begin{figure}[t]
    \centering
    \includegraphics[width=\linewidth]{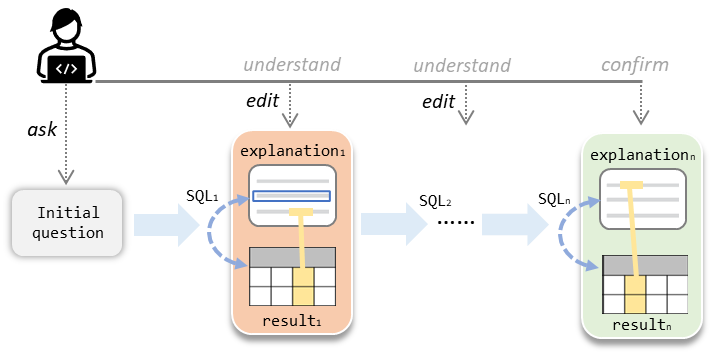}
    \caption{The iterative SQL refinement pipeline of {\tool}}  
    \Description{...}
    \label{fig:overview}
\end{figure}

However, most approaches only support feedback in constrained forms, e.g., answering multiple-choice questions~\cite{misp, dialsql, piia}, or changing keywords using a drop-down menu~\cite{diy}. Such constrained feedback is insufficient to fix complex errors in real-world tasks. 
Ning et al.~\cite{ning_empirical_2023} conducted a user study to evaluate three representative approaches for SQL generation and refinement, including MISP~\cite{misp}, DIY~\cite{diy}, and SQLVis~\cite{sqlvis}. They found no statistically significant difference in user performance compared to manually writing SQL queries.
Particularly, participants found it hard to understand the generated query and provide feedback.



To address this challenge, we draw inspiration from the grounding theory in communication~\cite{clark1991grounding}.
The theory suggests that effective communication requires a common ground, where speakers design utterances for listeners to understand and listeners provide feedback to resolve ambiguity and demonstrate understanding. 
However, recent studies in code generation have highlighted challenges due to insufficient communication between systems and developers~\cite{expectation_experience, taking_flight, barke2022grounded}. Systems often misinterpret the developer's intent, while developers often struggle to comprehend the generated code. 
This communication gap, which arises from a lack of common ground, results in code that does not align with user intent and hinders effective feedback~\cite{barke2022grounded}.

Based on this insight, we develop an interactive system, {\tool}, that leverages step-by-step \edit{SQL} explanations as the common ground between SQL generation models and users. Figure~\ref{fig:overview} provides an overview of the interaction pipeline. In each iteration, {\tool} generates an explanation \edit{in NL} to describe the individual steps in the generated SQL query. Through the rich interaction mechanisms provided by {\tool}, users can quickly navigate the explanation to understand the query and verify its behavior. If users recognize any erroneous steps, they can directly edit the explanation to inform the model which part of the SQL query should be regenerated and what the expected behavior is. 



\edit{Compared with existing techniques,} {\tool} \edit{has two key features, visual correspondence and intermediate query results.}
\edit{First, without an efficient way to navigate the database, users could easily become overwhelmed by the volume of data and complexity in the schema. Visual correspondence helps users instantly locate the related data by interacting with the entities mentioned in the explanation. They can also mentally connect elements mentioned in the explanation with elements in the database, which is helpful for sense-making.}
\edit{Second, the complex database schema makes certain query operations difficult to intuitively explain in NL. This is due to a logic gap between human understanding and database operations}~\cite{dbtalkback}. \edit{For instance, explaining a} \texttt{JOIN} \edit{operation to users based on primary and foreign keys can be challenging. Rendering intermediate results provides a convenient way for users to understand and verify the function of each step.}

\edit{Finally, we conducted a comprehensive evaluation of the usability of} {\tool}. \edit{This included two user studies with 38 participants in total and a quantitative experiment with 100 tasks. The first user study compared} {\tool} \edit{with MISP}~\cite{misp} \edit{and DIY}~\cite{diy}, \edit{demonstrating the effectiveness of our design choices over alternative designs. The second user study measured the contribution of each key feature in} {\tool}, \edit{showing that each feature significantly improves usability. The quantitative experiment shows the generalizability to a broad range of querying tasks. The results indicate that accuracy improves from 49}\% \edit{when no interaction is possible, to 89}\% \edit{when using} {\tool} \edit{and the user is familiar with it.}

\section{RELATED WORK}

\subsection{Interactive Support for Text-to-SQL}
\label{sec:related1}
There is a large body of literature on converting natural language (NL) questions to SQL queries, ranging from logic-based~\cite{logic1, logic2}, rule-based~\cite{rule1, SQLizer, ATHENA, Semantic-Tractability, construct_interface} to neural-based methods~\cite{smbop, ratsql, editsql, picard, sqlova}. However, these techniques only focus on improving the accuracy of text-to-SQL methods, instead of designing interactions to help non-experts understand and improve the query.


We summarize existing interactive support for text-to-SQL generation into two categories---(1) {\em explaining generated queries back to users} and (2) {\em soliciting user feedback to refine queries}. QueryVis~\cite{queryvis} and SQLVis~\cite{sqlvis} explain SQL queries by visualizing them as graphs. However, graphical representations can become unintuitive and overly complex for end-users~\cite{ning_empirical_2023}. Many existing systems resort to explanations \edit{in NL} instead~\cite{logos, explaininnl, SQL-to-text, diy, NaLIR}.  
For instance, Xu et al.~\cite{SQL-to-text} first convert an SQL query into a directed graph and then use a graph-to-sequence model to generate an NL summary of the query. DIY~\cite{diy} uses pre-defined templates to translate an SQL query into a step-by-step explanation \edit{in NL}. Similar to DIY~\cite{diy}, {\tool} also leverages step-by-step explanations \edit{in NL} but uses a different grammar-based method. The benefit is that such a grammar-based method can handle arbitrarily complex queries without being restricted to pre-defined templates. 
\edit{Furthermore, existing systems generally present explanations as static text, offering limited interactive capabilities for users to understand, validate, and refine queries. DIY}~\cite{diy} \edit{attempts to improve clarity by rendering intermediate results on a ``small-but-relevant'' sample database, but both their study}~\cite{diy} \edit{and the study by Ning et al.}~\cite{ning_empirical_2023} \edit{indicate that this approach can lead to user confusion. Specifically, users may find inconsistency between query results on the sampled database with the full database, as some relevant data may be missing.}
\edit{To address this issue}, {\tool} \edit{renders intermediate query results by executing on the entire database to make users fully comprehend the functionality of each step.}
\edit{Additionally, we propose further incorporating rich interactions in} {\tool} \edit{such as visual correspondence and direct query editing to augment the utility of explanations, thereby enhancing user engagement and understanding.}



A common way to solicit user feedback is through conversations~\cite{misp, dialsql, piia, nledit}. For instance, MISP~\cite{misp}, DialSQL~\cite{dialsql}, and PIIA~\cite{piia} detect a set of tokens with high uncertainty during the decoding process and ask multiple-choice clarification questions to users. 
In these systems, users can only passively clarify their intent by selecting from a limited set of options in the multiple-choice questions. NL-EDIT~\cite{nledit} allows users to proactively suggest SQL query edits via free-form text. 
Then, it uses an encoder-decoder model to convert the free-form text to a sequence of edits to refine the query. 
Despite the flexibility, incorporating such open-ended feedback is challenging. It requires the model to precisely infer which parts of the query to edit and which edits to apply. 

Direct manipulation~\cite{shneiderman1983direct} is an effective mechanism for rapid and accurate user feedback. Several text-to-SQL systems support direct manipulation and allow users to directly refine a query without knowing SQL syntax~\cite{diy, datatone, NaLIR, Eviza, Orko}. DIY~\cite{diy},  DataTone~\cite{datatone}, and NaLIR~\cite{NaLIR} allow users to directly change table names, column names, and values used in a query via a drop-down menu. In Eviza~\cite{Eviza} and Orko~\cite{Orko}, users can adjust a numeric value in a query using a slider. However, these systems only support a limited set of simple edits to SQL. They do not allow users to specify complex feedback, e.g., selecting data from two tables (i.e., \texttt{\textcolor[RGB]{172,41,0}{JOIN}}), grouping a set of data records (i.e., \texttt{\textcolor[RGB]{172,41,0}{GROUP BY}}), etc.

Our idea of grounding database queries with explanations resembles a recent work by Liu et al.~\cite{what_it_wants_me_to_say}. Liu et al.~ propose to use step-by-step explanations as a grounded abstraction to generate and refine Python code for spreadsheet data analysis. Despite the similarity in spirit, our work has several key differences in system design. First, like other text-to-SQL systems that \edit{explains SQL in NL}, Liu et al.~also only render the explanations as static, plain text. Compared with spreadsheet data analysis, database querying often involves complex data schema, multiple tables, and complex operations. Thus, our work presents richer interaction mechanisms to facilitate query comprehension and validation. Second, given user edits to a step-by-step explanation \edit{in NL}, {\tool} performs fine-grained query refinement at the clause or entity level, without the need to regenerate the entire query from scratch. By contrast, their system concatenates the explanations of individual steps as a new prompt and invokes CodeX~\cite{codex} to regenerate the entire Python code. This limits the utility of user feedback on step-by-step explanations and does not afford precise code refinement.

\vspace{-2mm}
\subsection{Human-AI Collaboration}
Promoting efficient collaboration between intelligent systems and humans has been a long-standing research topic in HCI.  This concept was first introduced in the seminal work on man-computer symbiosis~\cite{man-computer}. In that work, Licklider proposed that computers could perform routine tasks to pave the way for human insights, while human users could utilize their domain knowledge to make decisions that computers are not capable of making.
Nowadays, the inaccuracy of AI models in high-stake domains further necessitates collaboration between humans and AI. However, the lack of interpretability and communication convenience presents a significant challenge to effective human-AI collaboration~\cite{blackbox}. Even though humans have the potential to complement AI, they often struggle to understand AI’s states and effectively express their thoughts~\cite{liao2020questioning, bansal2021does, eiband2019people, kocielnik2019will, luger2016like, what_it_wants_me_to_say}. Specifically, if a user does not understand where the error is and what causes the error, they may find it difficult to provide effective instructions on fixing the error~\cite{dbtalkback}. 

Research from various domains has focused on explaining system behavior~\cite{flashPog, Rationalization, Generation-NL-Explanations, nl-explanation-visual-question, Tutorons}. 
\edit{For example, Head et al.'s work on Tutorons}~\cite{Tutorons} \edit{automatically generates context-relevant, on-demand in-situ explanations for code snippets, such as regular expressions, on web pages.} \edit{While Tutorons aims to bridge the gap between programmers and complex syntax,} {\tool} \edit{is designed for non-programmers to iteratively refine SQL queries in NL. Furthermore,} {\tool} \edit{makes these explanations editable in free-form NL, enabling intuitive user feedback.} {\tool} \edit{also incorporates visual correspondence and intermediate results to deepen user engagement compared to the static explanations provided by Tutorons.}

\section{USER NEEDS AND DESIGN RATIONALE}

\subsection{User Needs in SQL Generation}

To understand the needs of non-experts when querying databases, we conducted a literature review of previous papers that have done a formative study of text-to-SQL systems~\cite{sqlvis, diy}, have done a user study of existing tools~\cite{diy, ning_empirical_2023}, or have discussed the challenges and opportunities of text-to-SQL systems~\cite{dbtalkback, making-DB-usable, sqltutor, 3-important-determinants, Self-Assessment, visual-representations-in-science-education}.
Based on this review, we summarize three major user needs.

\begin{figure*}[t]
    \centering
    \includegraphics[width=0.85\linewidth]{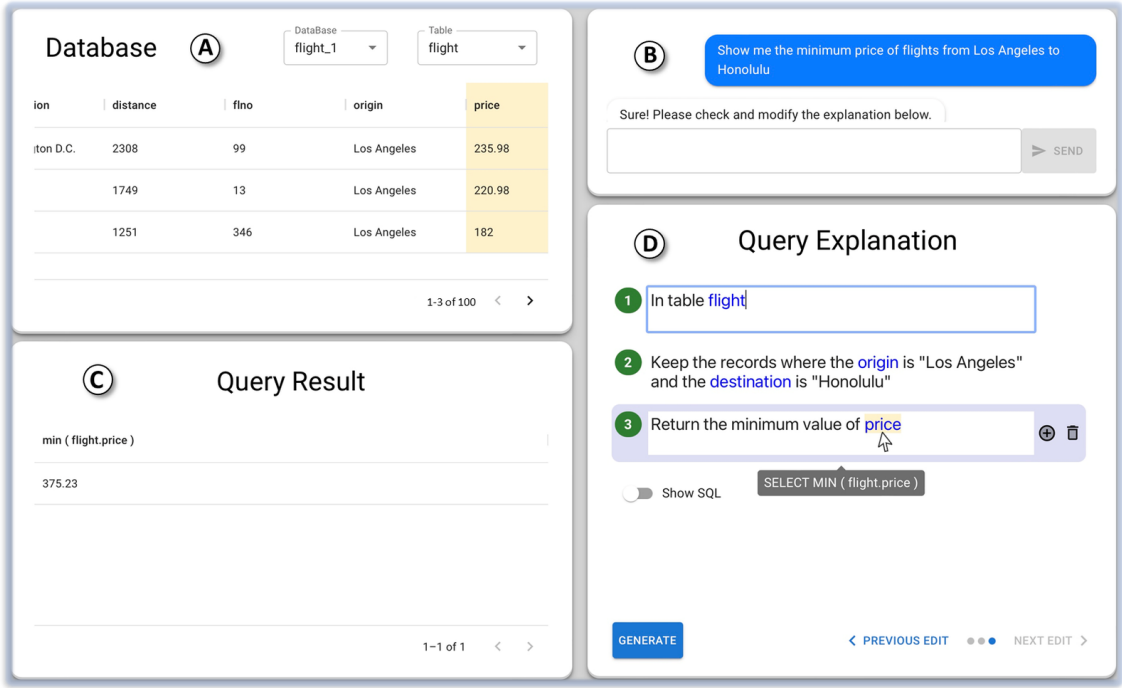}
    \caption{The user interface of {\tool}. (A) The \textit{Database} panel allows users to switch databases and tables in a database. It also allows users to manually inspect, search, and filter data. (B) The \textit{Question} panel allows users to ask a question to the database in natural language. (C) The \textit{Query Result} panel shows the query result as well as the intermediate result of individual steps when the user clicks each step. (D) The \textit{Query Explanation} panel renders the step-by-step \edit{SQL} explanation \edit{in natural language}. Users can directly edit the explanation to fix the incorrect behavior in a step, add new steps, or delete existing steps. 
    } 
    \label{fig:UI}
\end{figure*}

\textbf{\textit{N1: Users need effective methods to understand and validate a generated SQL query, so they can trust the result.}} 
Text-to-SQL systems are primarily designed for non-experts who are not familiar with SQL. Without additional support, the only way for them to validate the correctness of a generated query is to carefully examine if the query result looks reasonable. However, if a query involves too many rows, columns, and tables, it is cognitively-demanding and time-consuming to manually examine the query result. 
Kim et al.~\cite{dbtalkback} point out that for queries that return a large amount of data, it is useful for users to understand how the resulting data is retrieved from the database. So users can reason about the correctness of the query steps, rather than a large amount of resulting data. 
Jagadish et al.~\cite{making-DB-usable} argue that database systems can frustrate users if there is no explanation for some unexpected query results. In a user study with 12 participants, Narechania et al.~\cite{diy} found that participants appreciated explanations and wished to have multi-modal explanations to help them understand complex query operations, such as table joining and compound SQL clauses. 
Therefore, it is critical to help users validate the query behavior.

\textbf{\textit{N2: Users prefer SQL explanations that are concise, well-organized, and intuitive.}} 
Ning et al.~\cite{ning_empirical_2023} compared SQL explanations generated by three interactive systems in a user study. They found that the majority of participants preferred the shorter explanations provided by DIY~\cite{diy}, since they are easier to read and understand.
Leventidis et al.~\cite{queryvis} argued that the explanation \edit{in NL} can become very lengthy and verbose for complex SQL queries, limiting their readability and utility in practice. This is supported by a controlled lab study with 112 CS undergraduate students~\cite{3-important-determinants}. The study found that students can easily get lost when dealing with long and complex queries. When presenting a query in a more structured and succinct manner, students experienced significantly less cognitive load and performed much better in data query tasks. 
Therefore, we need to find the right level of abstraction that can concisely summarize the behavior of a SQL query in a clear and well-organized manner while matching user expertise.

\textbf{\textit{N3: Users need more flexible and expressive ways to provide feedback.}} Most existing systems support feedback in constrained forms, e.g., answering multiple-choice questions~\cite{misp, dialsql, piia}, changing incorrect keywords in a drop-down menu~\cite{diy}. 
This hinders users' ability to handle various SQL errors, especially for those requiring a completely new clause or subquery. 
A recent study~\cite{ning_empirical_2023} shows such mechanisms did not significantly improve the task completion rate or reduce the task completion time in complex text-to-SQL tasks compared with manually fixing a SQL query. 
Participants expressed frustration when they found they could not fix an error.
Thus, non-expert users need a more expressive and flexible way to guide the model to fix various SQL generation errors.

\vspace{-3mm}

\subsection{Design Rationale}

To support \textbf{N1}, we choose natural language as the communication vehicle, since it is understandable for non-experts and it is also flexible to express any kind of feedback.
An alternative design is to explain a query in a graphical representation~\cite{sqlvis, queryvis}. While graphs can be visually appealing, they can also become overly complex and counter-intuitive for non-experts~\cite{ning_empirical_2023}.

To support \textbf{N2}, we adopt step-by-step explanations and augment them with visual correspondence and intermediate query results.  
Users can utilize the visual correspondence to quickly locate relevant data and navigate a large database. This is particularly helpful when users are not familiar with the database schema or when there are many tables and columns. Displaying intermediate results helps users further validate the query behavior on concrete data and understand how each step contributes to the final result.

To support \textbf{N3}, {\tool} enables users to specify the correct behavior of a query step by directly editing the description of that step. There are several alternative designs for this feature. First, alternatively, we could ask users to rephrase the original NL question (i.e., {\em prompt engineering}) or provide NL feedback in a conversation~\cite{nledit}. However, recent studies show that prompt engineering is challenging and accurately interpreting NL feedback is as hard as interpreting the initial NL query~\cite{zamfirescu2023johnny, ning_empirical_2023, splash}. Some systems~\cite{diy, datatone, NaLIR} also enable users to provide feedback via direct manipulation. As discussed in Section~\ref{sec:related1}, these methods can only support limited edits. Another design option is to allow users to pinpoint errors in the intermediate or final query results. This design is effective for certain errors such as including extra columns and when the dataset is small. However, when the query results include many data records, it can be cumbersome and time-consuming to inspect all data and annotate which data records are wrong. 
Furthermore, regenerating the query based on input and output data may lead to overfitting, a known issue in programming-by-example techniques~\cite{padhi2019overfitting, lee2017towards}.

\vspace{-2mm}

\section{SYSTEM IMPLEMENTATION}
\label{sec:system}
\edit{In this section, we first describe the base text-to-SQL generation model used in}  {\tool} \edit{and a pilot study to understand the usability issues of the base model.}
\edit{We then detail the SQL generation process in} {\tool} \edit{and highlight three key features that facilitate efficient SQL query comprehension, validation, and repair.}

\vspace{-3mm}

\subsection{Base Model and Pilot Study}

\edit{The design of} {\tool} \edit{is model-agnostic.}
{\tool} \edit{is built upon a SQL generation system called STEPS}~\cite{emnlp-steps}, \edit{which provides algorithms and technical components to generate SQL explanations but only provides limited interaction support.}
\edit{We choose STEPS since it generates reliable grammar-based SQL explanation and provides a pre-trained text-to-clause model for SQL regeneration.}
\edit{However, one can replace STEPS with other models, e.g., using GPT-4 to generate SQL and step-by-step explanations.}

\edit{STEPS provides a limited primitive UI without careful consideration of the usability challenges. It only supports add or removing explanation steps.}
\edit{To better understand the usability challenges in STEPS, we conducted a pilot study with three participants. Each participant completed five data query tasks randomly selected from Spider}~\cite{spider}, \edit{with an average task completion time of 4 minutes. We subsequently interviewed participants about their experiences.}

\edit{The study revealed challenges in comprehending SQL through natural language descriptions alone. Participants found it time-consuming and cumbersome to manually navigate the database content, especially when explanations referenced multiple tables and columns. Participants often had to switch between tables and manually locate mentioned columns, making it difficult to validate query behavior.}
\edit{Participants also struggled to understand complex operations like JOINs and the concepts of primary and foreign keys, particularly in queries involving multiple tables. }
\edit{These findings highlighted the need for more intuitive ways to connect explanations with database entities and visualize query operations.}
\edit{It is important to ground the} \edit{SQL explanations on the data to help users better comprehend and validate the entities and operations mentioned in the explanation}. 
\edit{To address the usability issues,} {\tool} \edit{proposes rendering visual correspondence, as detailed in Section}~\ref{sec:visual}, \edit{and displaying intermediate query results, as detailed in Section}~\ref{sec:intermediate}.

\begin{figure*}[t]
    \centering
    \includegraphics[width=0.75\linewidth]{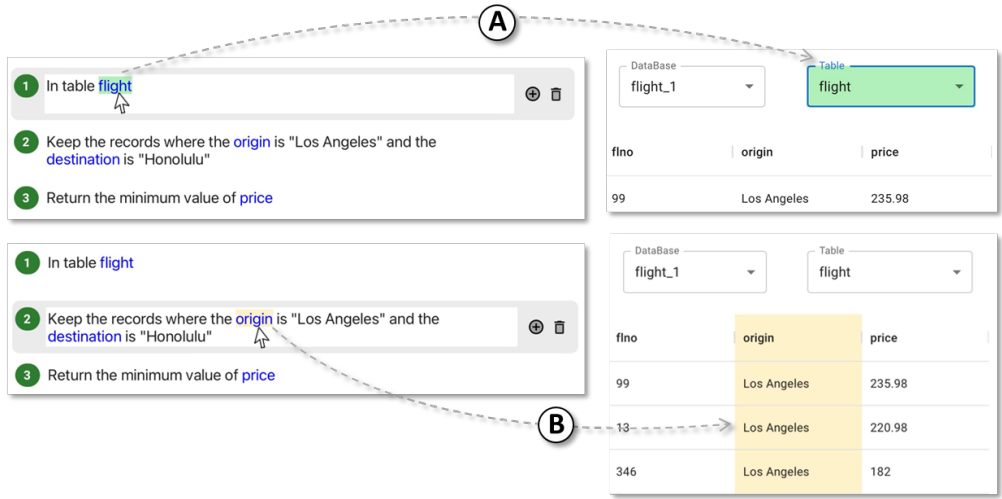}
    \vspace{1mm}
    \caption{(A) Hover over a {\em \bluetodo{table name}} and {\tool} automatically switches to the corresponding table and highlights it. 
        (B) Hover over a {\em \bluetodo{column name}} and {\tool} automatically highlights the entire column.
    } 
    \label{fig:highlight}
\end{figure*}

\subsection{SQL Query and Explanation Generation}


Given a SQL query generated by the underlying model, {\tool} generates a step-by-step explanation \edit{in NL} for the query. Following STEPS, we use the same text-to-SQL generation model, SmBoP~\cite{smbop}. Yet users can plug in any model they prefer to use. Furthermore, we adopt the same grammar-based explanation generation algorithm of STEPS~\cite{emnlp-steps}. 
The SQL query is decomposed into SQL clauses and each clause is then translated into NL descriptions based on SQL grammar.
\edit{This algorithm guarantees a deterministic and accurate translation from the SQL query to the NL description.}
Please refer to the STEPS paper~\cite{emnlp-steps} for technical details.

\edit{For users who know SQL, we still provide the option to view the generated SQL by clicking a toggle button below the explanation} (Figure~\ref{fig:explanation} \circled{d}).
\edit{This feature was requested by pilot study users who knew SQL and wished to double-check SQL code in our iterative design process. It is not designed for non-experts, since they are not familiar with SQL syntax and semantics.} {\tool}.
\edit{Reading NL descriptions and checking intermediate results is the main way for non-experts to validate SQL.}

\vspace{-3mm}

\subsection{Visual Correspondence via Highlighting}
\label{sec:visual}



As illustrated in Figure~\ref{fig:highlight}, {\tool} highlights the noun phrase of each database entity in blue in the \edit{SQL} explanation. 
We chose blue as it is the standard color for hyperlinks, which implies the highlighted entity can be interacted with.
When users hover over a highlighted entity, {\tool} will automatically navigate to the corresponding data in the database panel and highlight the corresponding data. Specifically, 
if the entity is a table, the drop-down menu turns green to indicate that the table is in focus (Figure~\ref{fig:highlight} \circled{A}). 
If the entity is a column, {\tool} automatically centers and highlights the column in yellow (Figure~\ref{fig:highlight} \circled{B}). 

A special case is nested SQL queries. The explanation generator of STEPS~\cite{emnlp-steps} splits a nested query into subqueries and generates an explanation for each of them separately. When Subquery A uses the result of Subquery B, the explanation of Subquery A will refer to the query result of Subquery B in natural language. To help users easily recognize which subquery's result is used by another subquery, {\tool} highlights the NL references with underscored hyperlinks. Section~\ref{sec:usage} illustrates this scenario.

We leverage the explanation generation method to establish the initial entity mappings. In particular, we instrument the explanation generator to log the translations between database entities and noun phrases in the explanation. {\tool} buffers these mappings in memory and dynamically highlights the database content when users hover over a noun phrase that maps to a database entity. 
Additionally, for nested SQL queries where one query may refer to the result of another, {\tool} uses natural language description (e.g., ``result of the first query'') to reference the result generated by a previous query. {\tool} also establishes a mapping and visually shows the correspondence between the reference and the previous result.
Since {\tool} allows for free editing of the explanation \edit{in NL}, users may rephrase some names, introduce new names, or even make typos. Whenever the explanation is edited, {\tool} re-calculates the mappings based on text similarity. 

\edit{We explored various approaches to calculate text similarity between database entities and SQL explanations. Initially, we considered using cosine similarity with word embeddings to better capture semantic relationships. However, we found that leveraging semantic information was often inaccurate when dealing with abbreviated or similar database entity names.}
\edit{Consequently, we opted for Levenshtein distance due to its optimal balance of precision and computational efficiency. Levenshtein distance can effectively capture nuances in spelling variations and similar names, while its lightweight nature ensures quick responsiveness in} {\tool}.

\subsection{Intermediate Query Results}
\label{sec:intermediate}
\begin{figure*}[t]
    \centering
    \includegraphics[width=0.85\linewidth]{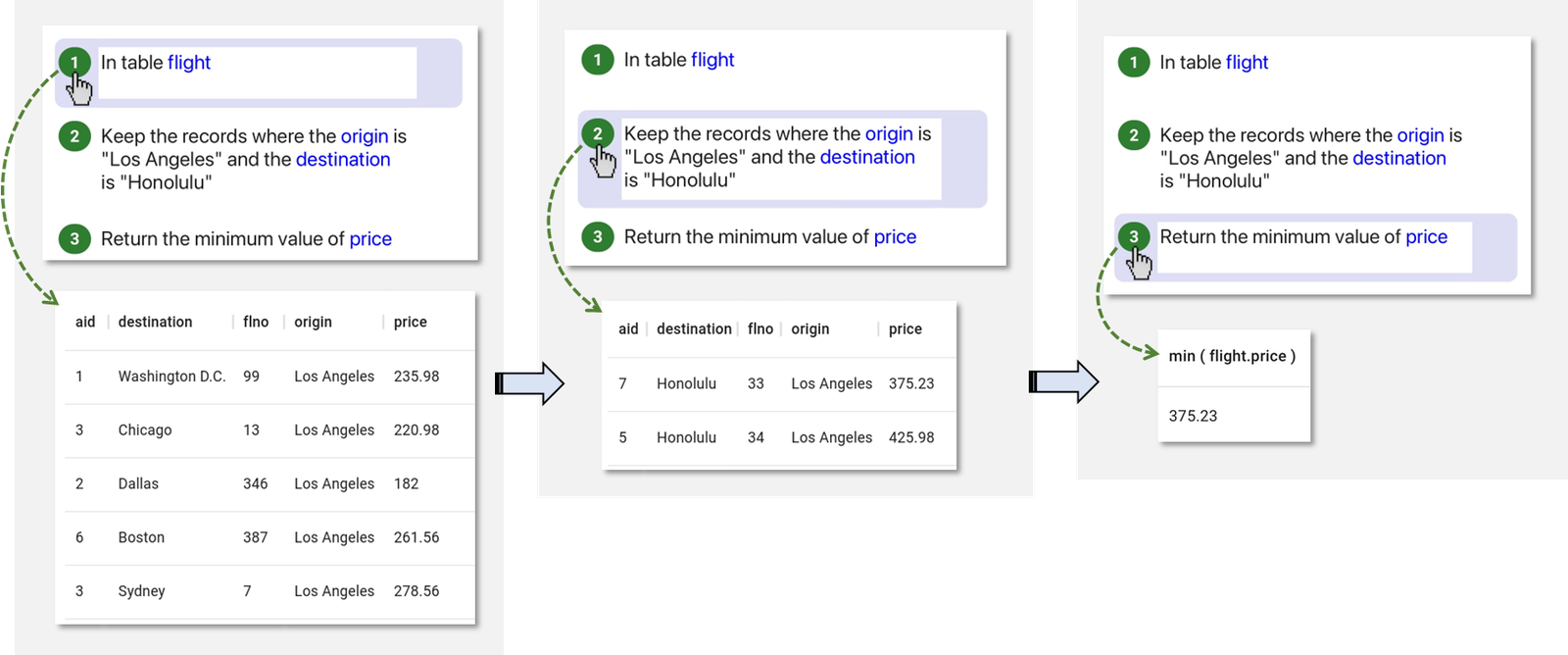}
    \caption{
        When clicking on the step number, the \textit{Query Result} view will visualize the intermediate result after the selected step.
    } 
    \label{fig:intermediate}
\end{figure*}

In order to help users understand the purpose of each step, {\tool} allows users to view the intermediate results corresponding to each step.
\edit{To compute the intermediate result,} {\tool} \edit{synthesizes a temporary SQL query by combining the current step with all preceding steps. However, simply concatenating SQL clauses from these steps may result in syntax errors or incomplete queries.}
\edit{For instance, missing the \texttt{SELECT} clause leads to an invalid SQL.}
\edit{To address this issue, we developed a synthesis algorithm inspired by the grammar-based explanation generation algorithm in STEPS}~\cite{emnlp-steps}. \edit{This algorithm converts a sequence of explanation steps back into a SQL query while following the grammar rules.}
\edit{Any missing clauses are automatically populated with dummy placeholders} (e.g., \texttt{\textcolor[RGB]{172,41,0}{SELECT} *}).
\edit{The resulting temporary SQL query is then executed on the database to compute the intermediate result.}

When the user clicks on the circled number of each step, the background of the corresponding step will turn blue, indicating this step is selected. The {\em Query Result} view (Figure~\ref{fig:UI} \circled{C}) is then updated to show the intermediate query result.

Figure~\ref{fig:intermediate} demonstrates an example.
When the user selects the first step, the database returns all the records in table \textit{flight}, with an initial temporary SQL query of ``\texttt{\textcolor[RGB]{172,41,0}{SELECT} * \textcolor[RGB]{172,41,0}{FROM} flight}''. When the user selects the second step, the database filters out all the records in the first step that do not satisfy this condition (i.e., flight from Los Angeles to Honolulu). 
The temporary SQL query becomes ``\texttt{\textcolor[RGB]{172,41,0}{SELECT} * \textcolor[RGB]{172,41,0}{FROM} flight \textcolor[RGB]{172,41,0}{WHERE} flight.origin = "Los Angeles" \textcolor[RGB]{172,41,0}{AND} flight.destination = "Honolulu"}''. 
When the user selects the third step, the database returns the minimal price from the remaining records.
The temporary SQL query becomes ``\texttt{\textcolor[RGB]{172,41,0}{SELECT MIN} (flight.price) \textcolor[RGB]{172,41,0}{FROM} flight \textcolor[RGB]{172,41,0}{WHERE} flight.origin = "Los Angeles" \textcolor[RGB]{172,41,0}{AND} flight.destination = "Honolulu"}''. 

\vspace{-2mm}
\subsection{Query Refinement by Explanation Editing}
\label{sec:fix}

\begin{figure}[t]
    \centering
    \includegraphics[width=0.975\linewidth]{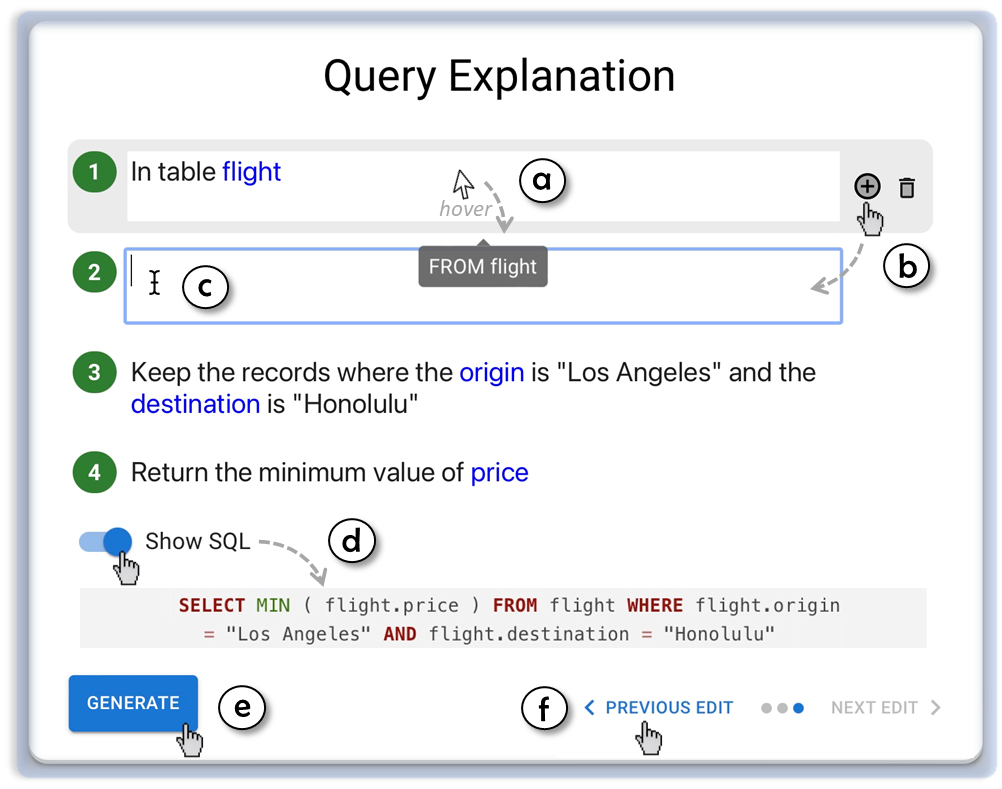}
    \caption{Interacting with the step-by-step explanation 
        }
    \label{fig:explanation}
\end{figure}

While inspecting the explanation \edit{in NL} and the intermediate query results, if a user finds an erroneous step, they can directly edit the description of that step to specify the correct behavior (Figure~\ref{fig:explanation} \circled{c}).  
Users can type in any description in free-form text, without being confined to a certain format. 
Users can also add a new step at any position or delete any existing step by clicking on the ``Add'' or ``Delete'' button next to an existing step (Figure~\ref{fig:explanation} \circled{b}).
Once the user has finished modifying the explanation, they can click the ``Generate'' button to request {\tool} to regenerate the SQL based on the edited explanation (Figure~\ref{fig:explanation} \circled{e}). 
\edit{A complex SQL query can sometimes consist of multiple subqueries (a SQL statement with only 1 SELECT keyword) concatenated together with set operations (e.g. UNION).}
\edit{Within a single subquery, the position of a newly added step is not important in our design, as} {\tool} \edit{can reorder and rectify all steps based on clause types to form a valid subquery. However, for a complex query involving multiple subqueries, users should ensure that new steps are added to the explanation of the corresponding subquery.} 
Finally, {\tool} allows users to check their edit history and undo/redo some edits by clicking on the stepper buttons at the bottom (Figure~\ref{fig:explanation} \circled{f}).

To interpret the edited explanation and correct the error, we adopt the same text-to-clause model used in STEPS~\cite{emnlp-steps}. It achieves an exact match accuracy of 90.6\%.
Like the text-to-SQL model, this model is independent of our system and can easily be replaced by other models.
After regenerating a clause, {\tool} merges it with the original query and automatically rectifies any syntax errors or conflicts.
Please refer to the STEPS paper for more technical details.

\vspace{-2mm}

\section{USAGE SCENARIO}
\label{sec:usage}
Suppose Alice is a social scientist and she wants to investigate the correlation between people's mobility behavior patterns and flight prices since the pandemic. 
She needs to analyze a large database with millions of flight records distributed in many different tables. 
As the first step, she wants to know the airport with the most flights to the most popular destination in the first quarter of 2022.
Alice finds it time-consuming to manually filter the database and find the desired data record. 
Furthermore, since the information spans across multiple tables, Alice does not know how to filter the data based on multiple conditions on multiple tables simultaneously.

Therefore, Alice decides to try {\tool}.
She asks, ``\textit{Show me the airport which has the most flights to the most popular destination in the first quarter of 2022.}''  in the {\em Question} panel.
Then, based on Alice's question, {\tool} automatically generates a SQL query and executes it in the database.
Alice looks at the query result but she is not sure whether it is correct.
Therefore, Alice chooses to read the step-by-step explanation of the SQL query in the \textit{Query Explanation} panel. Since the generated query is a nested query, {\tool} explains the inner query as the first and the outer query as the second query below:

\vspace{1mm}
\noindent\colorbox[rgb]{0.95,0.95,0.95}{Start the first query}
\begin{enumerate}[topsep=2pt]
  \item Merge data in table \textcolor{blue}{flight} and table \textcolor{blue}{travel}.
  \item Keep the records where \textcolor{blue}{month} is January.
  \item Split the data into groups based on the \textcolor{blue}{destination}.
  \item Sort the groups based on the number of records in descending order, and return the first record.
  \item Return the \textcolor{blue}{destination}.
\end{enumerate}

\vspace{0.5mm}
\noindent\colorbox[rgb]{0.95,0.95,0.95}{Start the second query}
\begin{enumerate}[topsep=2pt]
  \item In table \textcolor{blue}{travel}.
  \item Keep the records where the \textcolor{blue}{destination} is \textcolor{blue}{\underline{the result of the} \underline{first query}}.
  \item Split the data into groups based on the \textcolor{blue}{airport code}.
  \item Sort the groups based on the number of records in descending order, and return the first record.
  \item Return the \textcolor{blue}{airport name}.
\end{enumerate}


The step-by-step \edit{SQL} explanation gives Alice a high-level understanding of the generated SQL query.
She roughly understands the purpose of the first query is to find the most popular destination, and the purpose of the second query is to find the airport with the most flights to this destination (hyperlinked blue text in Step 2).

However, Alice is unsure about what kind of code is associated with ``\textit{airport code}'' in Step 3 of the second query. 
Thus, she wants to see some actual data in the database.
However, when she tries to locate the related data in the database, she notices there are many tables and some tables even include hundreds of columns. 
She does not want to do this manually.
Instead, Alice hovers her mouse over the highlighted text ``\textit{airport code}'' in this step.
The database panel automatically switches to the table that includes ``\textit{airport code}'', centering and highlighting data in this column in yellow.
After reviewing the data in the database, Alice confirms that the airport code is a unique identifier used when booking a flight. She is confident this step has no issue.

Alice is curious about the difference between Table ``\textit{flight}'' and Table ``\textit{travel}'' in Step 1 of the first query, so she hovers the mouse over these two entities respectively.
By moving the mouse between them, the corresponding tables are being switched accordingly.
Alice clearly notices why these two tables need to be merged. This is because ``month'' is stored in the Table ``flight'', while ``destination'' is stored in Table ``travel''.

Since Alice is not familiar with SQL, she still does not understand how these two tables are merged in this step. 
Therefore, she clicks on this step to view the intermediate result.
The intermediate result shows a combined table with columns from the ``\textit{flight}'' table and the ``\textit{travel}'' table. Alice checked a few data records in the merged table and compared them with the original records in these two tables to confirm that they were indeed consistent.

As Alice reads individual steps in the \edit{SQL} explanation, she notices that Step 2 of the first query is wrong. 
It seems {\tool} misinterpreted the meaning of ``the first quarter'' as ``January'', and it also ignored the year constraint. 
Instead of rephrasing her original question, Alice modifies the description of Step 2 in the first query by explicitly specifying the beginning and ending months.
She then adds a new step below to instruct the system to only consider data in Year 2022.  
Below is the modified \edit{SQL} explanation. 

\vspace{1mm}
\noindent\colorbox[rgb]{0.95,0.95,0.95}{Start the first query}
\begin{enumerate}[topsep=2pt]
  \item[(1)] Merge data in table \textcolor{blue}{flight} and table \textcolor{blue}{travel}.
  \item[(2)] Keep the records where \textcolor{blue}{month} is \newline [January] $\rightarrow$ [\textbf{between January and March}]. \textcolor[RGB]{190,190,190}{(Updated)}
  \item[+ (3)] \fbox{Make sure the year in 2022.} \textcolor[RGB]{190,190,190}{(Added)}
  \item[(4)] Split the data into groups based on the \textcolor{blue}{destination}. 
  \item[(5)] Sort the groups based on the number of records in descending order, and return the first record.
  \item[(6)] Return the \textcolor{blue}{destination}.
\end{enumerate}

\vspace{0.5mm}
\noindent\colorbox[rgb]{0.95,0.95,0.95}{Start the second query}
\begin{enumerate}[topsep=2pt]
  \item In table \textcolor{blue}{travel}.
  \item Keep the records where the \textcolor{blue}{destination} is \textcolor{blue}{\underline{the result of the} \underline{first query}}.
  \item Split the data into groups based on the \textcolor{blue}{airport code}.
  \item Sort the groups based on the number of records in descending order, and return the first record.
  \item Return the \textcolor{blue}{airport name}.
\end{enumerate}

Then Alice clicks on the \textit{Generate} button to update the query.
She receives a new airport name and a new \edit{SQL} explanation.
By checking the explanation and the intermediate results again, Alice is convinced that the result is correct and exactly what she needs.

\section{USER STUDY I: Comparison with Other Interactive Approaches}

To investigate the usability of the holistic system, we conducted a within-subjects user study with {\participants} participants in comparison to two representative interactive systems, MISP~\cite{misp} and DIY~\cite{diy}.
To ensure a fair comparison, we have redesigned the front-end user interfaces of MISP and DIY following the same design style as {\tool} (detailed in Appendix~\ref{app:ui}). Furthermore, we have replaced the original SQL generation model in MISP and DIY with the same SQL generation model used in {\tool}. 
In this way, we normalize the impact of the visual appearance and also the underlying models on user performance in the comparison. 

\subsection{Participants}
We recruited participants through the mailing lists in an R1 university. To investigate the impact of user expertise on {\tool}, we selected participants with three different levels of familiarity with SQL. In total, we recruited {\participants} participants. 15 of them had never heard about or used SQL before ({\em end-user}); 10 knew the basics of SQL but had to search online to recall syntax details ({\em novice}); 5 could fluently write SQL queries ({\em expert}). 14 participants were undergraduate students, 4 were master's students, and 12 were PhD students. We shared the consent form in the recruitment email and obtained their consent before each study. Each participant received a \$25 gift card as compensation for their time.

\subsection{Comparison Baselines}

MISP~\cite{misp} and DIY~\cite{diy} are two state-of-the-art interactive approaches for SQL generation. They adopt two typical mechanisms, question-answering and direct manipulation.

\textbf{MISP} \edit{uses a question-answering interaction mechanism, where users clarify ambiguities through multiple-choice questions. To enable fair comparison, we created a graphical interface for MISP similar to} {\tool}, \edit{excluding the} \textit{Query Explanation} view (Figure~\ref{fig:UI} \circled{D}), \edit{and used the same text-to-SQL model}~\cite{smbop} as {\tool}.

\textbf{DIY} \edit{employs direct manipulation, allowing users to correct mappings between SQL tokens and natural language phrases using drop-down menus. We adapted the replication from Ning et al.} \cite{ning_empirical_2023} \edit{with a} {\tool}\edit{-like interface and the same underlying model} \cite{smbop}.

Appendix~\ref{app:ui} provides details and screenshots of baseline UIs.

\subsection{Tasks}
\label{sec:tasks}

We performed stratified random sampling on a widely used text-to-SQL benchmark, Spider~\cite{spider}, to create a pool of 48 tasks. This task pool includes 12 easy tasks, 12 medium tasks, 12 hard tasks, and 12 extra hard tasks, according to the difficulty classification from Spider. Table~\ref{tab:tasks3} in the Appendix show 12 representative tasks.

\subsection{Protocol}
\label{sec:protocol}
Each study consisted of three sessions, one for each tool. 
We randomized the order of assigned tools to mitigate learning effects. 
Each session starts with participants watching a tutorial video about the assigned tool.
Then participants were given several minutes to practice and get familiar with the tool before working on real tasks. 
Once they were done practicing, participants were given 10 minutes to complete 8 assigned SQL tasks using the designated tool. Specifically, we selected 2 tasks per difficulty level from the pool of 48 tasks.  We randomized the order of the 8 tasks in each session to counterbalance the impact of task difficulty levels (e.g., doing easy tasks first vs.~doing difficult tasks first). If a participant found a task too difficult to solve, they were allowed to skip it.
For each task, participants were asked to first read the task description and then ask an initial natural language question to the assigned tool. After receiving the generated query and the query result, the participant could further validate and correct the generated query using the interaction mechanisms provided by the tool.  

At the end of each session, participants were asked to complete a post-task survey to share their experiences. The survey included the NASA Task Load Index (TLX) questions~\cite{NASA-TLX} and several 7-point Likert-scale questions to rate their perception of the assigned tool. 
After all sessions, participants completed a final survey, in which they directly compared all the tools and shared their overall thoughts about the usefulness of the tools. We recorded each study with the permission of the participants. Participation took 79 minutes overall on average.

\subsection{User Performance}
\label{sec:performance}

\begin{figure}[htb]
    \centering
    \includegraphics[width=0.945\linewidth]{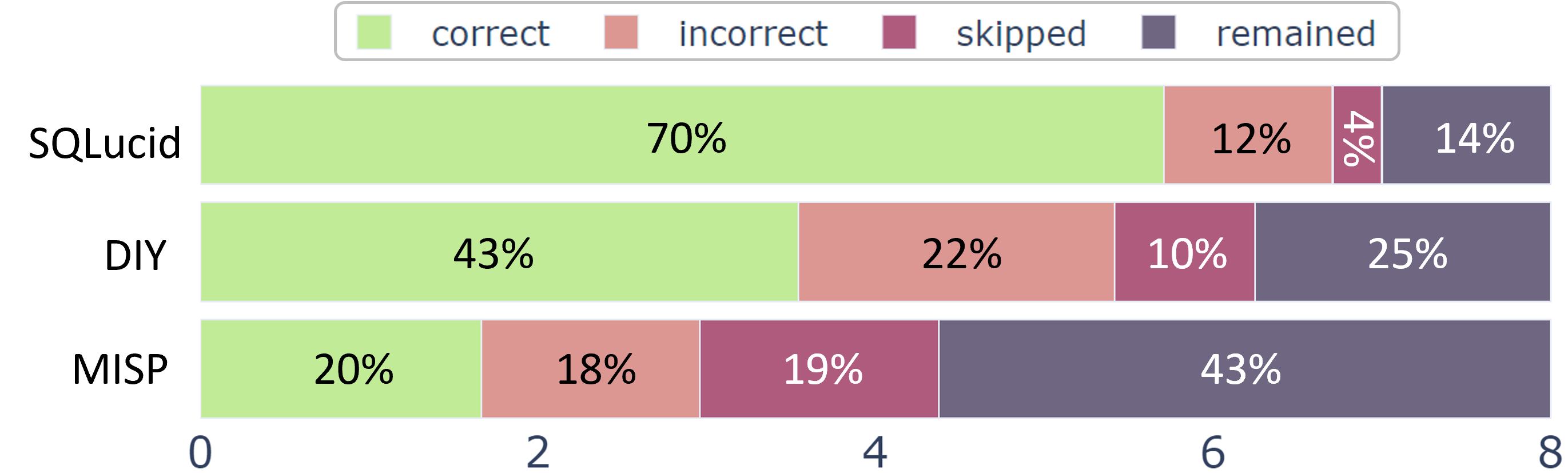}
    \vspace{-1mm}
    \caption{Distribution of correctly completed, incorrectly completed, skipped, remaining tasks  (Study 1)}
    \label{fig:performance}
\end{figure}

\begin{table}[htb]
    \centering
    \caption{Task Completion Accuracy (Study 1).}
    \resizebox{0.75\linewidth}{!}{%
    \begin{tabular}{lcc}
        \toprule
              & \textbf{Task completion accuracy} & \textbf{SD} \\
        \midrule
        MISP~\cite{misp}  & 56\%  & 30\% \\
        DIY~\cite{diy}   & 67\%   & 20\% \\
        {\tool} & \textbf{85\%}   & 13\% \\
        \bottomrule
    \end{tabular}
    }
    
    \label{tab:task_accuracy}
\end{table}

\noindent\textbf{Task Completion Rate.}
Figure~\ref{fig:performance} shows the distribution of completed, correct, skipped, and remaining (i.e., tasks that were not even tried due to the time limit) tasks when using different tools. 
An ANOVA test showed that the mean differences among the number of completed tasks, correct completion, skipped tasks, and remaining tasks when using different tools are all statistically significant ($p$-value = 1.75e-16, 7.99e-25, 7.62e-6, 1.21e-2 respectively).

Specifically, participants using {\tool} completed 6.6 out of 8 tasks, while participants using MISP and DIY completed 3.0 and 5.4 tasks respectively.
This result suggests that {\tool} can accelerate the speed of task completion.
Furthermore, when using {\tool}, participants skipped only 4\% of the tasks, compared with 10\% when using DIY and 19\% when using MISP. This implies that {\tool} can provide more effective support to help participants make progress on challenging tasks, leading to fewer skipped tasks.

To measure the correctness of completed tasks, we calculate the \textit{task completion accuracy}---the number of correctly completed tasks divided by all completed tasks, excluding skipped tasks and remaining tasks. 
Table~\ref{tab:task_accuracy} shows the result. 
Participants using {\tool} also achieved the highest task completion accuracy, 85\%. In contrast, participants using MISP and DIY only achieved 56\% and 67\% accuracy, respectively. In other words, in 44\% and 33\% completed tasks, participants using MISP and DIY thought they had arrived at a correct query when in fact, the query was still wrong. 
These results imply that {\tool} can significantly improve user productivity when querying databases and help them effectively recognize query errors and generate correct queries with high accuracy.

\vspace{1mm}
\noindent\textbf{Utility Rates of Different Features.}  
To better understand the utility of different features, we analyzed recordings and gathered utility rates of features. 
For each task, participants intentionally navigated data by checking the visual correspondence 10.2 times.
Participants rendered the intermediate results 3.5 times.
In 48\% of assigned tasks, {\tool} generated the correct query in the first iteration and participants did not edit the \edit{SQL} explanation.
In 47\% of assigned tasks, {\tool} generated a wrong query in the first iteration, and it took 1.8 edits to fix. 
In 5\% of assigned tasks, participants either rephrased the question or skipped the task. 

These values show that participants heavily rely on visual correspondence and the intermediate results to understand the query. 
With these features, participants can quickly identify and successfully fix errors with only a few edits per task, improving the initial query generation accuracy from 48\% to 85\%.
We also analyzed the recordings of participants using DIY and MISP. We found that DIY and MISP generated the initial query correctly in 48\% and 51\% of the assigned tasks. However, due to the limitation of their interaction methods, participants could not effectively understand the generated query and only fixed a limited number of queries, resulting in a 56\% and 67\% final accuracy, respectively. 



\vspace{2mm} 

\noindent{\bf The Impact of User Expertise.} Table~\ref{tab:expertise} shows the number of correctly completed tasks for participants with different levels of expertise. Overall, compared with MISP and DIY, {\tool} consistently improved the task completion correctness and efficiency across all levels of SQL expertise. Specifically, the performance gap between different expertise levels when using {\tool} is narrow. An ANOVA test showed that when using {\tool}, there is no statistically significant difference in the number of correctly completed tasks between different levels of SQL expertise ($p$-value=0.88). This implies that {\tool} can help bridge the expertise gap among users when querying databases. 





\begin{table}[htb]
    \centering
    \caption{Correctly completed tasks by expertise level}
    \begin{minipage}{0.9\linewidth}
        \centering

        \resizebox{\linewidth}{!}{
            \begin{tblr}{
                cells = {l},
                row{1} = {c},
                cell{1}{2} = {c=2}{},
                cell{1}{4} = {c=2}{},
                cell{1}{6} = {c=2}{},
                vline{2} = {1}{},
                vline{2,4,6} = {2-5}{},
                hline{1,6} = {-}{0.08em},
                hline{2} = {2-7}{},
                hline{3} = {-}{},
                rowsep = 1pt, 
            }
                  & \textbf{MISP~\cite{misp}} &      & \textbf{DIY~\cite{diy}}  &      & \textbf{{\tool}}  &      \\
                  & \textbf{\#Corr.}  & \textbf{SD}   & \textbf{\#Corr.} & \textbf{SD}   & \textbf{\#Corr.} & \textbf{SD}   \\
            End-User & 1.6 & 0.91 & 3.3 & 1.13 & 5.4 & 0.91 \\
            Novice & 1.4 & 0.67   & 3.5 & 1.27   & 5.7 & 1.06   \\
            Expert & 2.2 & 1.10   & 3.8 & 1.30   & 5.9 & 1.22
            \end{tblr}
        }
        \label{tab:expertise}
    \end{minipage}
\end{table}

\vspace{-3mm}

\noindent{\bf The Impact of Task Difficulty Levels.} 
Table~\ref{tab:difficulty} shows the number of correctly completed tasks at different levels of difficulty when using different tools. Overall, compared with MISP and DIY, {\tool} consistently improved the task completion correctness and efficiency across all levels of task difficulty.
In particular, {\tool} significantly improves user performance on hard and extra-hard tasks. Compared with using MISP, participants using {\tool} completed almost 9X and 3X more extra-hard tasks correctly compared with using MISP and DIY. P10 wrote, ``\textit{I really enjoyed [{\tool}] a lot better than the previous two. I can use it to answer complex questions. Sometimes the system made a mistake at the first step, but I can easily correct it or add more constraints.}''

\begin{table}[htb]
    \centering
    \begin{minipage}{0.9\linewidth}
        \centering
        \caption{Correctly completed tasks by difficulty level}
        \resizebox{\linewidth}{!}{
            \begin{tblr}{
                cells = {l},
                row{1} = {c},
                cell{1}{2} = {c=2}{},
                cell{1}{4} = {c=2}{},
                cell{1}{6} = {c=2}{},
                vline{2} = {1}{},
                vline{2,4,6} = {2-6}{},
                hline{1,7} = {-}{0.08em},
                hline{2} = {2-7}{},
                hline{3} = {-}{},
                rowsep = 1pt, 
            }
                  & \textbf{MISP~\cite{misp}} &      & \textbf{DIY~\cite{diy}}  &      & \textbf{{\tool}}  &      \\
                  & \textbf{\#Corr.}  & \textbf{SD}   & \textbf{\#Corr.} & \textbf{SD}   & \textbf{\#Corr.} & \textbf{SD}   \\
            Easy    & 0.81 & 0.74     & 1.40 & 0.69  & 1.62 &  0.50 \\
            Medium  & 0.49 & 0.62   & 1.19 & 0.72   & 1.48 & 0.60   \\
            Hard    & 0.21 & 0.51   & 0.54 & 0.61   & 1.39 & 0.58  \\
            Extra hard  & 0.12 & 0.31   & 0.35 & 0.63   & 1.14 & 0.66  
            \end{tblr}
        } 
        \label{tab:difficulty}
    \end{minipage}

\end{table}

\subsection{User Confidence and Cognitive Load}
\begin{figure}[htb]
    \centering
    \includegraphics[width=\linewidth]{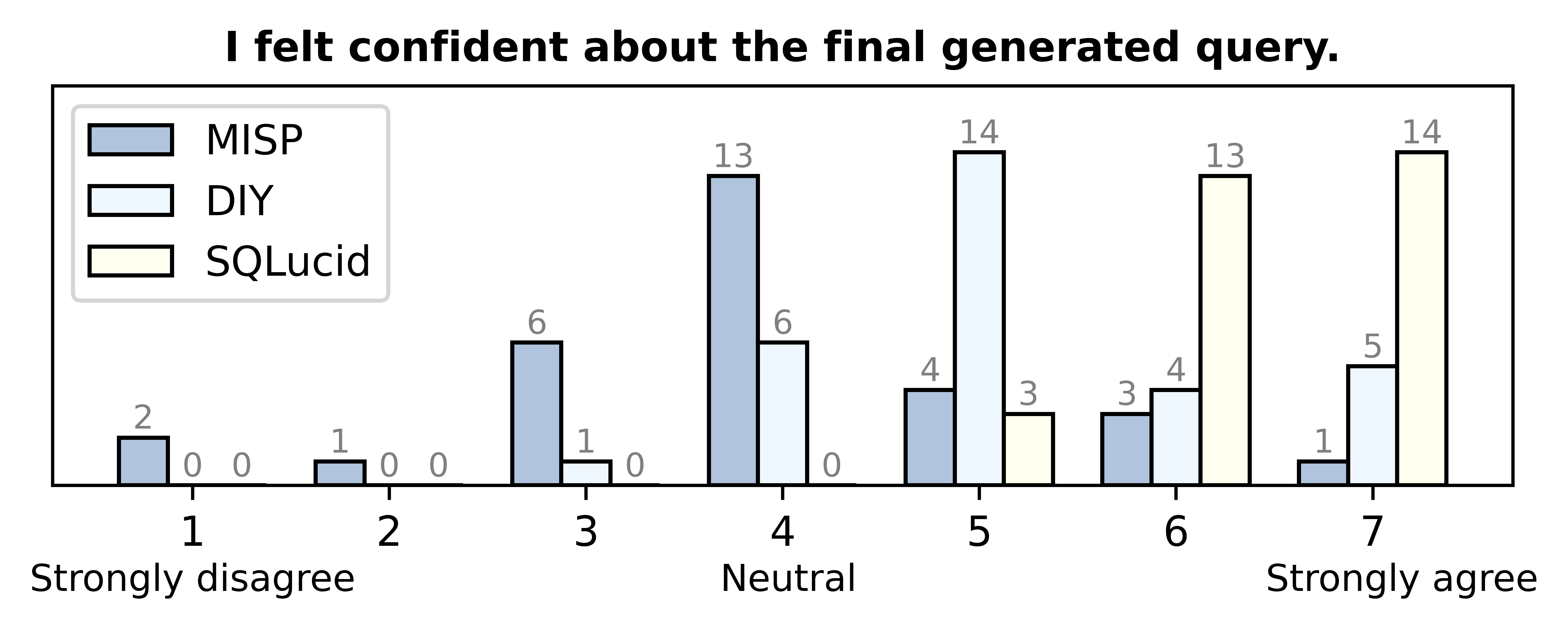}
    \caption{User Confidence Ratings (Study 1)}
    \label{fig:confidence}
\end{figure}

In the post-task survey, participants self-reported their confidence about generated queries when using different tools on a 7-point scale. 
Figure~\ref{fig:confidence} shows the distribution of users' confidence levels. The average confidence level is 6.42 when using {\tool}, compared with 3.79 and 5.29 when using MISP and DIY. An ANOVA test showed that the mean differences are statistically significant ($p$-value = 1.53e-11). 
Based on a qualitative analysis of user responses, we believe this improvement was largely attributed to the visual correspondence and intermediate features provided by {\tool}. 
P23 wrote, ``\textit{I felt most confident using {\tool} because it provided the most information on how a natural language query was interpreted and carried out. For example, I could see intermediate results and explanations of steps in natural language, allowing me to easily gauge whether the process was correct or not.}''
P9 reported, ``\textit{When I was trying to explore the data for the other two tools, it was a bit challenging. But with related tables and data highlighted w.r.t. the explanations made it easier to navigate the data.}''

\begin{figure}[htb]
    \centering
    \begin{minipage}{\linewidth}
        \includegraphics[width=\linewidth]{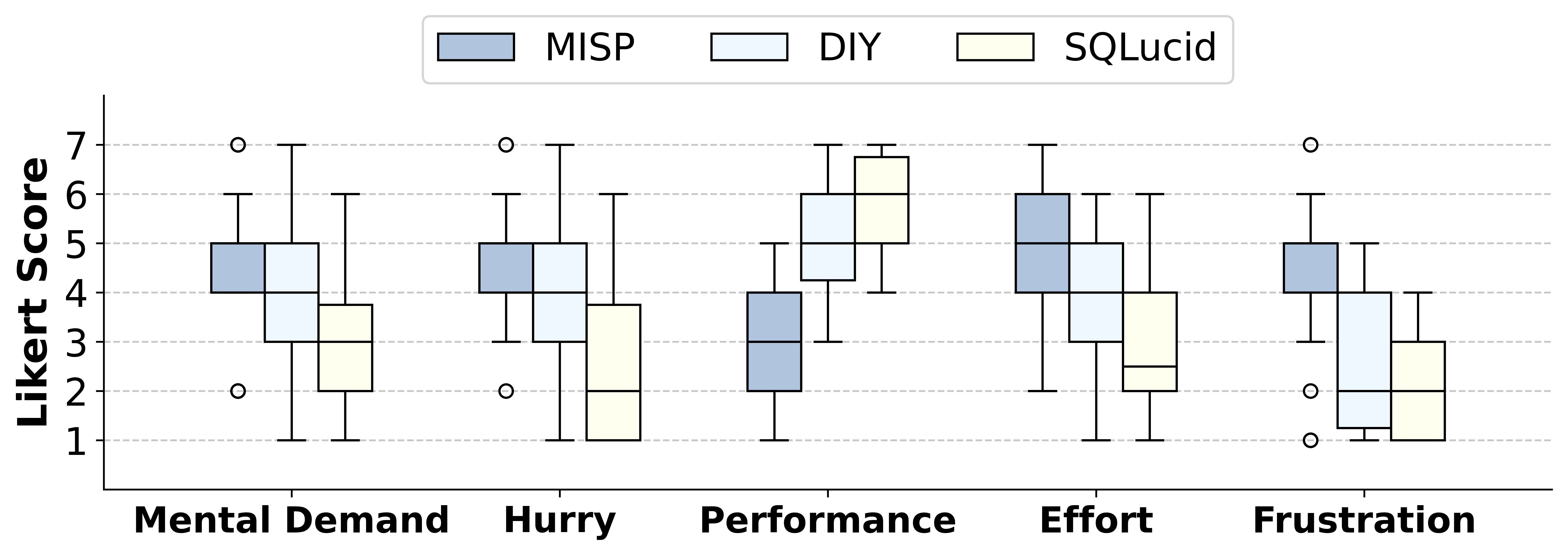}
        \caption{NASA Task Load Index Ratings (Study 1)}
        \label{fig:NASA}
    \end{minipage}
\end{figure}

Figure~\ref{fig:NASA} shows participants’ ratings on the five cognitive load factors from the NASA TLX questionnaire~\cite{NASA-TLX}. The ANOVA test demonstrates that the mean differences are all statistically significant ($p$-value=8.26e-4, 7.83e-06, 6.04e-13, 2.57e-06, 8.10e-08 respectively). 
The result confirms that {\tool} can reduce users' cognitive load by creating interactive \edit{SQL} explanations with visual correspondence and intermediate results, which serves as a common ground between users and the database.
P19 wrote a comprehensive comment to illustrate the convenience provided by {\tool}---``\textit{{\tool} helps me query the database and debug my query completely with natural language, which is good because I do not know SQL. The intermediate results help me locate bugs easily, so I don't need to debug my entire query. The natural language interpreter is so flexible that I do not need to change my writing style to accommodate it. All inferences are performed on the database level, so I don't need to specify which table I should look into. The highlight feature also helps me navigate the database.}''

\subsection{User Ratings of Individual Features}

\edit{In the post-task survey, we prepared six 7-point Likert scale questions for participants to rate the usefulness of key features in} {\tool}.
\edit{The most appreciated features were} {\em being able to understand the SQL query via the step-by-step} \edit{explanation and} {\em being able to directly edit the explanation in natural language to fix an error}. 
\edit{Other features in} {\tool} \edit{ere also appreciated by the majority of participants. More discussion is detailed in Appendix}~\ref{app:individual_features}.

\subsection{User Preference and Feedback}
When asked about the tool they preferred to use for their real-world data query needs, all 30 participants selected {\tool}. We coded participants' responses in the post-study survey and identified two main reasons why they liked {\tool} more. 
First, 27 participants mentioned that the explanations provided by {\tool} were more understandable and useful. 
Particularly, the visual correspondence and intermediate result features bring more interactivity in {\tool}, and greatly enhance users' ability to identify errors. 

Second, 21 participants pointed out that {\tool} is the most useful among all conditions because the direct editing of \edit{SQL} explanations \edit{in NL} is more convenient and requires less effort. 
P23 wrote, ``\textit{{\tool} was the most usable because I felt that it was very easy and fast to correct mistakes in interpretation using this tool. 
For example, I could directly use language to edit some of the intermediate steps to get the correct order of steps. I think this is fast and convenient.
}''



In the post-task survey, we also asked participants what additional features may help them better solve the task.
Seven participants 
mentioned that it would be helpful to see confidence scores associated with each step, because they can pay more attention to those steps with lower confidence.
P1 wrote, ``\textit{I wish to see a confidence score that indicates if I need to check or debug something. }''
Furthermore, three participants mentioned they would like to see some suggestions when editing the \edit{SQL} explanation.
P11 suggested that ``\textit{providing suggested expressions may diminish the chances for the normal language question to be misinterpreted.}''
Finally, two participants mentioned that it might be useful for {\tool} to generate multiple answers and let the user choose one.

\section{USER STUDY II: Ablation Study of Key Features in {\tool}}
\label{app:user_study_2}
To investigate the effectiveness of each feature in {\tool}, we conducted another within-subjects user study with 8 participants, comparing {\tool} with three of its variants.

\subsection{Participants, Baselines, Tasks and Protocol}
We followed the same procedure as Study 1 to recruit 8 participants for this study. 
4 of them had never heard about or used SQL before ({\em end-user}); 2 knew the basics of SQL but had to search online to recall details of the syntax when writing a SQL query ({\em novice}); 2 could fluently write SQL queries ({\em expert}).

We created three different variants of {\tool} as comparison baselines by ablating the two key features: (1) no visual correspondence, (2) no intermediate results, (3) no visual correspondence \& no intermediate results (i.e., Text SQL explanation only). 

In this study, we used the same tasks (Section~\ref{sec:tasks}) and followed the same protocol (Section~\ref{sec:protocol}) as the first user study. On average, each study took about 61 minutes in total.

\begin{figure}[h]
    \centering
    \includegraphics[width=0.945\linewidth]{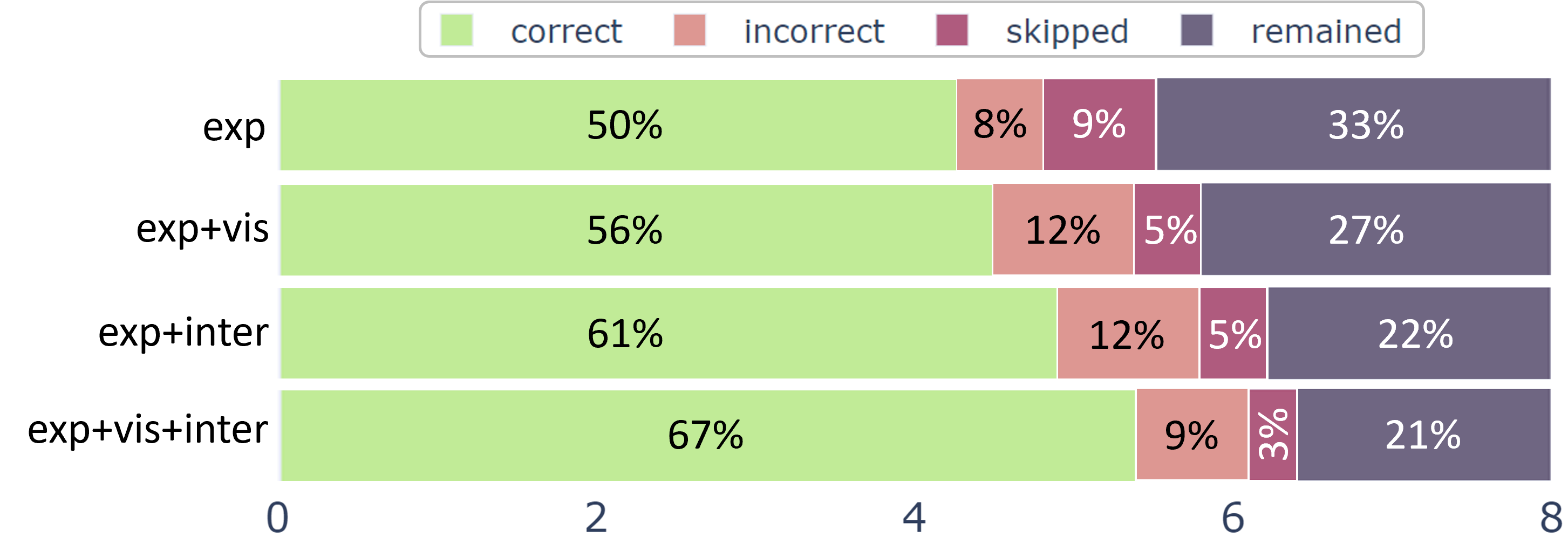}
    \vspace{-1mm}
    \caption{Distribution of correctly completed, incorrectly completed, skipped, remaining tasks when using different versions of {\tool} (Study 2)}
    \label{fig:performance_variant}
\end{figure}

\vspace{-2mm}

\subsection{User Performance}
Figure~\ref{fig:performance_variant} shows the distribution of completed tasks, correct tasks, skipped tasks, and remaining tasks.
Table~\ref{tab:accuracy_variant} shows the task completion accuracy similar to user study 1.
An ANOVA test showed that the mean differences among these values are statistically significant, except for skipped tasks ($p$-value = 2.12e-02, 3.36e-02, 3.3e-01, 1.03e-03, 1.21e-2 respectively).


Specifically, when the \edit{SQL} explanation is plain text, participants completed 4.9 out of 8 tasks with a completion accuracy of 81.6\%, and skipped 0.75 out of 8 tasks.
When the \textit{visual correspondence} feature is activated, participants completed 5.5 tasks with a completion accuracy of 83.1\% and skipped 0.375 tasks.
When the \textit{intermediate query result} feature is activated, participants completed 5.9 tasks with a completion accuracy of 83.5\% and skipped 0.375 tasks.
When both features were activated, participants completed 6.4 tasks with a completion accuracy of 84.3\% and skipped 0.25 tasks.
The result implies both the two features can reduce task completion time and increase user performance.


\begin{table}[htb]
    \centering
    \caption{Task Completion Accuracy (Study 2).}
    \resizebox{0.9\linewidth}{!}{%
    \begin{tabular}{lcc}
        \toprule
              & \textbf{Task completion accuracy} & \textbf{SD} \\
        \midrule
        Text Explanation Only  & 81.6\%  & 7.9\% \\
        +Visual   & 83.1\%   & 16.3\% \\
        +Intermediate & 83.5\%   & 11.9\% \\
        +Visual+Intermediate & \textbf{84.3\%}   & 8\% \\
        \bottomrule
    \end{tabular}%
    }
    \label{tab:accuracy_variant}
\end{table}

\begin{figure}[h]
    \centering
    \includegraphics[width=\linewidth]{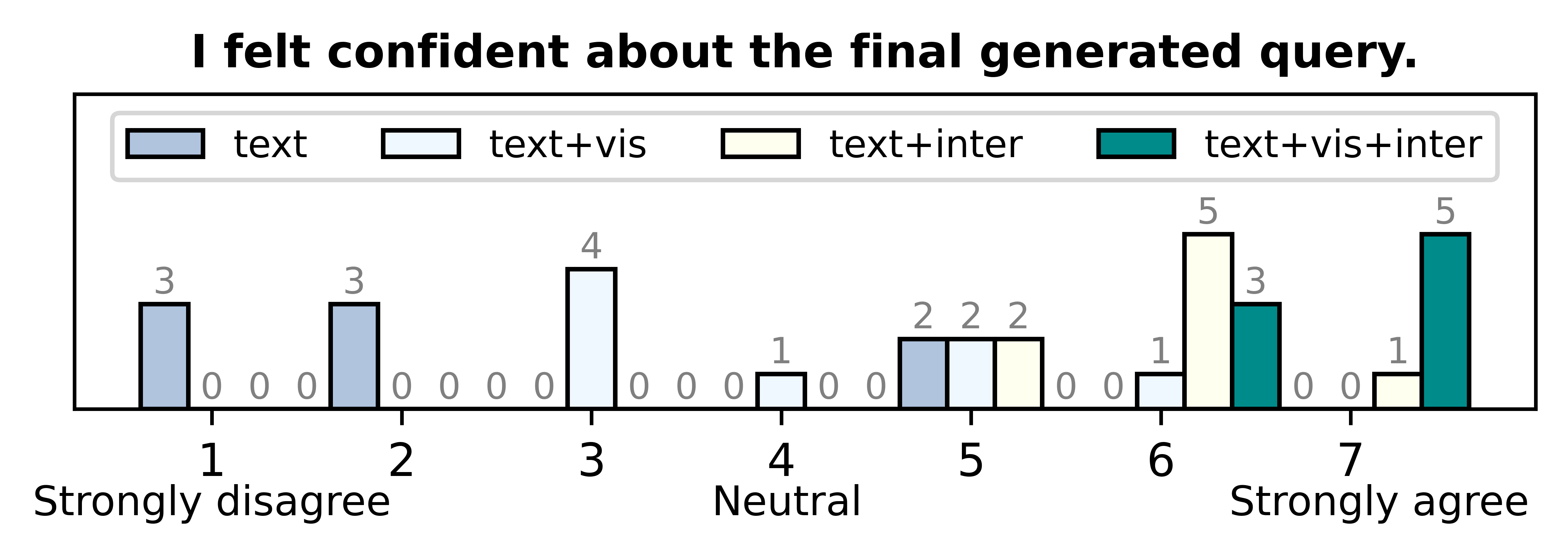}
    \caption{User Confidence Ratings (Study 2)}
    \label{fig:user_confidence2}
\end{figure}

\begin{figure}[h]
    \centering

    \begin{minipage}{\linewidth}
        \includegraphics[width=\linewidth]{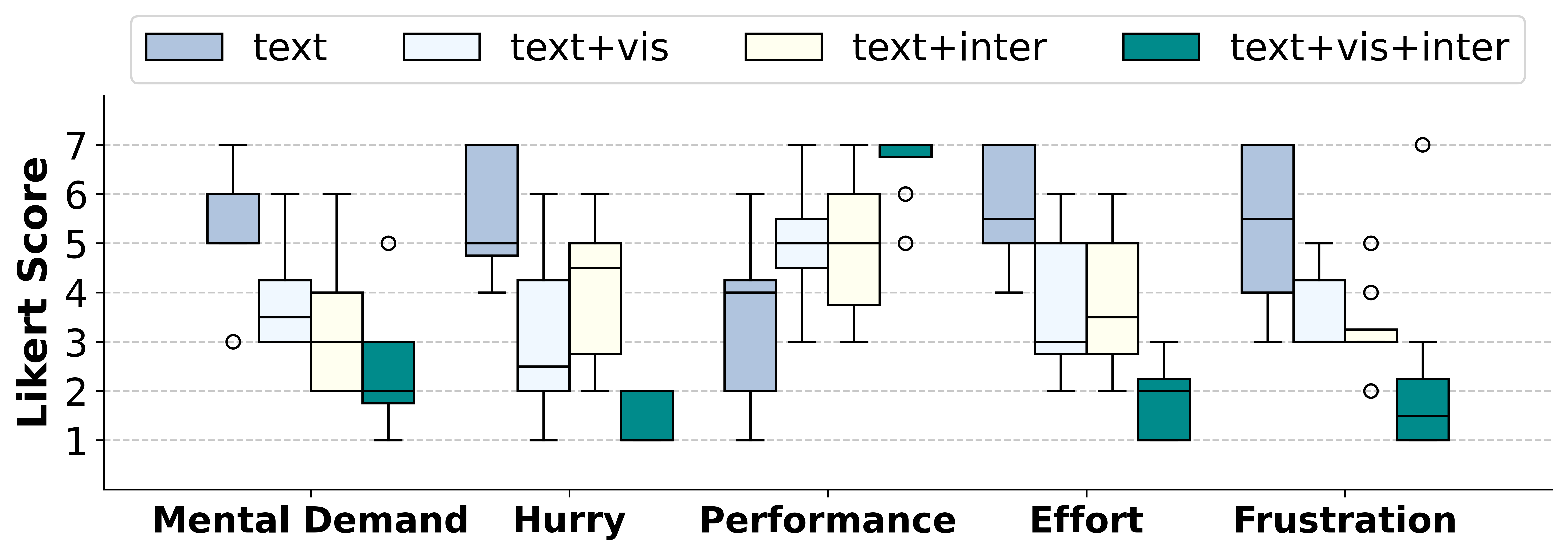}
        \caption{NASA Task Load Index Ratings (Study 2)}
        \label{fig:NASA2}
    \end{minipage}

\end{figure}

\subsection{User Confidence and Cognitive Load}

Figure~\ref{fig:user_confidence2} shows the participants' confidence with different tools. An ANOVA test shows that the mean differences across different conditions are statistically significant ($p$-value=1.32e-12). Figure~\ref{fig:NASA2} shows user ratings on the five cognitive load factors from the NASA TLX questionnaire. An ANOVA test shows that the mean differences in all five dimensions are statistically significant ($p$-value=1.65e-05, 2.71e-06, 1.14e-16, 7.34e-07, 2.53e-09 respectively).
Participants using {\tool} with all features activated have the lowest cognitive load and highest confidence.
The result shows both the \textit{visual correspondence} feature and the \textit{intermediate query result} feature serve as great supplements to the \edit{plain SQL} explanations.

We analyzed the post-study survey responses and found that these two features contributed to different aspects of user performance. 
Specifically, the \textit{visual correspondence} feature aids in data navigation, thereby saving more time. 
P1 wrote, ``\textit{[Without visual correspondence,] I need to use the scrolling bar a lot. That is annoying and tedious.}''
On the other hand,
the \textit{intermediate query result} feature focuses on improving user comprehension of the explanation, which brings more confidence.
P4 wrote, ``\textit{Intermediate results give me confidence about the final outcome.}''
Additionally, this feature provides information that users may not have asked for, but can offer additional context, thereby reducing their cognitive load.
P1 commented, ``\textit{Intermediate steps can help me check back and forth based on my needs. Without this feature, I only get a piece of information. If I want to know more, I need to ask multiple times.}''

Overall, features in {\tool} complement each other and collaboratively enhance the interactivity of \edit{SQL} explanations.
P2 made a comprehensive comment about the variant with only plain textual explanation, ``\textit{Without these features, my interest in using this system decreases a lot, because I need to find the data by my eyes and the mouse. 
Although it explains the procedure in English and provides the final result, I can't see the relationship between the sentences and the real data. 
Without seeing the relationship, it might be correct, but I question my understanding and do not trust it. Besides, sometimes when my request is too complex for the system to handle, I don’t know which step is wrong.}''

\section{Quantitative Evaluation}

\edit{To evaluate the generalizability of} {\tool} \edit{, we further conducted a quantitative experiment where the first author completed 100 database query tasks. The results of this study can be interpreted as the upper bound of user performance of} {\tool}.

We followed the same sampling strategy in the user studies, including 25 easy tasks, 25 medium tasks, 25 hard tasks, and 25 extra hard tasks from Spider~\cite{spider}.
For each task, the first author examined the database, read the natural language description, and tried to solve the task using {\tool}. The task was considered completed when a correct query result was obtained.
\edit{This simulates an ideal condition where a user is familiar with the tool and has sufficient knowledge of database queries.} This case study is to investigate to what extent {\tool} can solve query tasks, regardless of its learnability or usability.


The experimenter completed all 100 tasks with an accuracy of 89\%, in an average of 1.9 minutes (median=0.9, SD=0.6) for each task. Table~\ref{tab:case_study} shows the task completion accuracy at different levels of task difficulty.
5 out of 100 tasks were failed due to user misunderstanding. For example, while the correct SQL for \textit{"French citizens"} should be \texttt{``Citizenship = France''} according to the data in the database, the query produced by the experimenter had \texttt{``Citizenship = French''}.
It is possible that additional affordances that proactively provide information about matches with content in the database could address these issues.
4 out of 100 tasks failed to be completed due to the complex query structure, e.g., a query with multiple subqueries.
The experimenter decided to skip them because they were time-consuming to solve. 





\begin{table}[htb]
\centering
\caption{Task completion at different levels of task difficulty}
\resizebox{0.9\linewidth}{!}{
\begin{tabular}{l|cccc|c} 
\toprule
 & \textbf{Easy} & \textbf{Medium} & \textbf{Hard} & \textbf{Extra hard} & \textbf{Overall} \\
\midrule
\textbf{Accuracy} & 96\% & 96\% & 76\% & 88\% & 89\% \\
\bottomrule
\end{tabular}
}
\vspace{1mm}
\label{tab:case_study}
\end{table}


\section{DISCUSSION}


\subsection{Design Implications}

Based on the evaluation results, we found that the primary enabler for {\tool} lies in the bi-directional, natural language (NL) communication channel it establishes between human users and SQL generation models.
Compared to directly editing and refining the original question (i.e., prompt engineering), editing the step-by-step explanations provides a more structured way to give feedback and allows users to pinpoint the error. Furthermore, by breaking down a lengthy explanation into shorter descriptions of individual steps, {\tool} can clearly and systematically explain the behavior of a query. The editability of these explanations allows human users to identify the specific step where an error occurs and directly propose a correction by altering the NL description of the erroneous step. This design enables users to offer more precise feedback \edit{and incrementally build a complex query} than they could by providing high-level suggestions in a multi-turn dialogue (e.g., MISP~\cite{misp}, \edit{ChatGPT}), thereby streamlining the SQL regeneration process. 

The success of {\tool} also echoes the {\em grounding} theory in communication~\cite{clark1991grounding}. Grounding theory states that conversation is a collaborative effort aimed at establishing common ground or shared knowledge. In interactions with intelligent systems, such as SQL generation models, the system should offer evidence of understanding in response to a user's input, enabling the user to assess progress toward their goal. In our work, the editable step-by-step explanation serves as the common ground for communication between an SQL generation model and a human user---{\em the model explains a generated query step by step, while the human user corrects the model's misinterpretation by directly editing the explanation.}
\edited{Furthermore, both the visual correspondence and the intermediate query result features further enhance the grounding.}

Our work further illustrates that comprehending system behavior and repairing system breakdowns are highly interdependent activities. This is in line with previous studies of conversational agents~\cite{ashktorab2019resilient, beneteau2019communication}, which argue that users must first understand the current state of the system and the cause of a breakdown to choose an effective repair strategy. By providing a detailed explanation with intermediate results, {\tool} enables users to rapidly grasp the query's behavior and identify the root cause of an incorrect query result. This helps users to efficiently pinpoint the erroneous part of the query and give accurate and effective suggestions to fix it. 
Additionally, this design offers users greater flexibility in expressing their intent and feedback compared to relying on constrained mechanisms to gather feedback~\cite{misp, diy, binary, construct_interface, piia}.



\subsection{Using Interactive Explanation for Task Decomposition}

Task decomposition is a long-standing challenge in program synthesis and code generation~\cite{jayagopal2022exploring, vaithilingam2022expectation, lee2017towards, gulwani2015inductive}. Several approaches support task decomposition by asking users to specify intermediate steps~\cite{kandel2011wrangler, yessenov2013colorful, hu2021assuage}. For instance, Wranger~\cite{kandel2011wrangler} recommends a ranked list of operators at each synthesis step and asks users to select which operator to use and fill in the parameters. Using such systems requires users to be familiar with the underlying programming language and also actively think about intermediate steps to arrive at the final solution. Prior work shows that non-experts often find it difficult to decompose a complex task into sub-tasks~\cite{lee2017towards}. 

The editable step-by-step explanation can serve as a scaffold to guide non-experts to decompose a complex task. Compared with prior work, {\tool} does not require users to actively make a task decomposition plan. Instead, the step-by-step explanation can be viewed as an initial decomposition plan proposed by {\tool}. Users only need to read and correct it. In particular, the step-by-step structure of the explanation will spontaneously inspire users to think about the intermediate steps and make it easier to recognize incorrect or missing steps. Since the explanation is communicated in natural language, users also do not need to know the semantics of the underlying programming language.

As we were developing this system, the rise of Large Language Models (LLMs) has brought another possibility 
 for task decomposition. Recent studies have shown that LLMs are capable of breaking down a large task into smaller subtasks with proper instructions~\cite{patel2022question, COT, COT-self-consistency, jiang2023self, song2022llm}. For instance, Chain-of-Thought (CoT) Prompting~\cite{COT} allows users to provide several examples of how to solve a problem analytically step by step and leverages the in-context learning capability of LLMs to decompose similar problems. Given a natural language query, one can use CoT to decompose it and generate a step-by-step plan with basic query operations. However, one caveat is that LLMs may hallucinate and generate an incoherent plan with non-sensical steps, as shown by many studies~\cite{LLM_errors, LLM_COT_inconsistency}. In contrast, our grammar-based explanation method is strictly grounded in the SQL components and provides a faithful representation of computation steps in a query. We also provide a dedicated method to incorporate user refinement on individual steps to fix query generation errors.


\subsection{Application to Other Domains}

We believe that \edited{our interface design} can be generalized to adjacent domains, such as enabling user validation and repair in code generation~\cite{chen2021evaluating, hu2021assuage}, data transformation synthesis~\cite{drosos2020wrex, flashPog}, web automation~\cite{chasins2018rousillon, leshed2008coscripter}, smartphone app automation~\cite{li_sugilite:_2017, li_pumice:_2019}, and regular expression synthesis~\cite{zhang2020interactive}.
Programs in these domains can be naturally decomposed into smaller components (e.g., program statements, API calls) and then explained in natural language in a similar step-by-step fashion. However, for certain domains such as tensor transformation synthesis~\cite{zhou2022intent}, step-by-step explanations may not be the most suitable approach, as code in these areas often involves complex concepts and computation steps, such as linear algebra, which are challenging to clearly explain in natural language.

\subsection{SQL Experts vs.~Non-Expert Users}

{\tool} \edit{is specifically designed for non-experts who need to interact with databases but lack SQL expertise.} 
\edit{Reading NL descriptions and checking intermediate results is the main way for non-experts to validate SQL queries.}
Our analysis of user performance across varying SQL expertise levels reveals that the performance gap between end-users, novices, and experts has been substantially reduced when utilizing {\tool}. 
Our user study results show \edit{adding and removing NL steps are intuitive for non-experts. Users can freely edit the NL description of a query step and} {\tool} \edit{updating the corresponding SQL component accordingly based on a text-to-clause model.}
\edit{If one step (e.g., group students into clusters by years) is missing in a query (e.g., compute the average GPA of students for each year), it is easy to recognize it from the NL description and the results.}

While our focus was on non-experts, we discovered that {\tool} can also enhance the productivity of SQL experts. \edit{For complex tasks that necessitate joining multiple tables or creating compound queries,} {\tool} \edit{offers a solid starting point from which SQL experts can iteratively and incrementally refine the query.}
\edit{For example, users can build two simple subqueries respectively and reference one within the other to form a more complex query.}


Another unintended benefit was that participants in our study found {\tool} to be valuable for learning SQL. Five participants who were unfamiliar with SQL actively reported that their ability to read basic SQL queries improved as a result of using {\tool}, and they expressed a desire to continue using it for practical SQL learning. Participant P12 commented, ``\textit{It was nice to see the generated SQL code with human language. I believe I could learn SQL using this tool.}'' Similarly, P24 stated, ``{\em I wish {\tool} can be made available as a website. It can be used to teach beginners SQL knowledge and I believe they are willing to pay for it.}''

\subsection{Limitation and Future Directions}
There are several limitations in the design of our user study. First, although our participants represented a wide range of expertise levels in SQL, they were all university students. In the future, we plan to recruit industrial practitioners to study the real-world adoption and ecological validity of {\tool}. We will also conduct semi-structured interviews and surveys to gather feedback from industrial practitioners. 
Second, \edit{we did not explicitly measure user perception of accuracy, but user confidence is a useful proxy for it. Figure}~\ref{fig:confidence} \edit{shows a significant improvement in the confidence of} {\tool} \edit{compared to DIY and MISP.}
\edit{Figure}~\ref{fig:user_confidence2} \edit{shows each key feature in} {\tool} \edit{contributes to increase user confidence.}



The current design of {\tool} offers room for further improvement.
\edit{First, to further enhance its educational potential,} {\tool} \edit{can establish a triple-linkage among the SQL statement, SQL explanation, and corresponding database content. Combined with the intermediate query results, this can serve as a promising learning tool for SQL beginners to understand both the syntax and semantics of SQL queries.}
Furthermore, as suggested by several participants, {\tool} can benefit from displaying more information about the SQL generation process, such as model confidence scores. This additional information could direct users' attention and help them determine which steps of the query they should prioritize.
\edit{Another future direction could focus on automatically reordering edited steps.} {\tool} \edit{currently assumes users know exactly where to add new steps. Supporting automatic step reordering can eliminate this assumption.}



\section{CONCLUSION}
\edited{
This paper presented {\tool}, a novel interactive SQL refinement interface that enables users to effectively query data from relational databases using natural language. 
{\tool} integrates editable explanation, visual correspondence, intermediate query results, and other auxiliary features.
These features echoed with each other, creating a grounded natural language interface with rich interactions for users to understand the generated queries, identify errors, and correct any errors.
A user study with 30 participants shows that {\tool} can help users query data more quickly and accurately, with increased confidence and reduced cognitive load. 
A user study with 8 participants demonstrates the effectiveness of key features in {\tool}. A quantitative experiment with 100 query tasks indicates that {\tool} can be generalized to various tasks.
}



\begin{acks}
We thank anonymous reviewers for their helpful and detailed feedback, as well as the time and care they spent reviewing our work.
We thank all the participants in the pilot study and two user studies for their valuable comments.
This work was supported in part by Amazon Research Award and the National Science Foundation (NSF Grant ITE-2333736).
\end{acks}

\bibliographystyle{ACM-Reference-Format}
\bibliography{sample}

\appendix





\section{User Ratings of Individual Features}
\label{app:individual_features}

\begin{figure*}[htb]
    \centering
    \includegraphics[width=\linewidth]{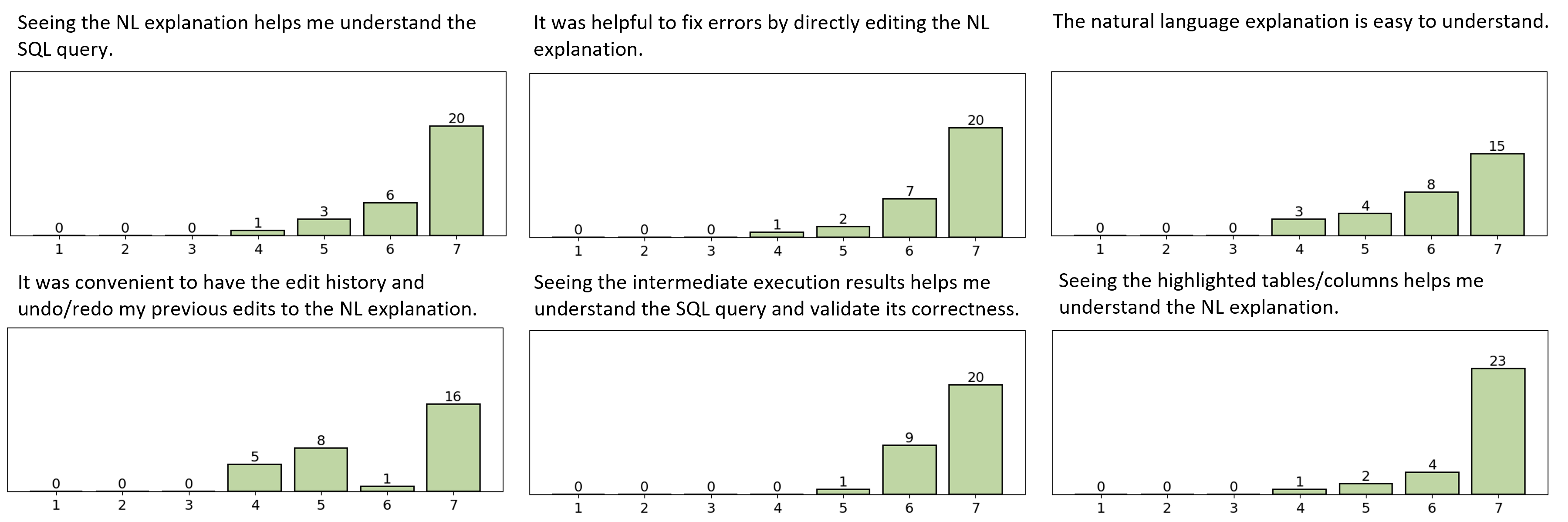}
    \caption{User ratings on individual features (1---strong disagreement, 7---strong agreement)}
    \label{fig:features}
\end{figure*}

In the post-task survey, participants rated the usefulness of key features of {\tool} in 7-point Likert scale questions. Figure~\ref{fig:features} summarizes the distribution of user ratings. 

We found that the majority of participants were satisfied with each feature in
{\tool}. The most appreciated features were {\em being able to understand the SQL query via the step-by-step explanation} and {\em being able to directly edit the explanation in natural language to fix an error}. P10 wrote, ``\textit{I really enjoyed this tool [{\tool}] a lot better than the previous two. Doing everything in natural language is way more direct. I don't have to answer strange questions or click confusing options [in drop-down menus]...By editing the steps, I was able to get more answers than previous tools.}''

Furthermore, 30 participants agreed or strongly agreed that ``\textit{seeing the intermediate execution results helps me understand the SQL query and validate its correctness.}'' 
P23 commented, ``\textit{I liked how intermediate steps and results were shown so users could see how the system interpreted the query.}''
29 participants agreed or strongly agreed that ``\textit{seeing the highlighted tables/columns helps me understand the NL description.}''
P5 wrote, ``\textit{the highlighting feature is useful for users to locate the corresponding elements quickly.}''
Even the least appreciated feature---the edit history of \edit{SQL} explanations---was still considered convenient by the majority of participants (25/30). 
P14 wrote, ``\textit{I also liked how easy it was to go in and edit the query as well as go back if I made a mistake.}''

\section{User study tasks}
\begin{table*}
\centering
\caption{Some example tasks in the user study}
\label{tab:tasks3}
\resizebox{0.92\textwidth}{!}{
\begin{tblr}{
  column{1} = {c},
  cell{2}{1} = {r=3}{},
  cell{5}{1} = {r=3}{},
  cell{8}{1} = {r=3}{},
  cell{11}{1} = {r=3}{},
  vline{2} = {2,3,4,5,6,7,8,9,10,11,12,13,14,15,16}{},
  hline{1-2} = {-}{},
  hline{5} = {-}{},
  hline{8} = {-}{},
  hline{11} = {-}{},
  hline{14} = {-}{},
  hline{3-14} = {2-3}{},
}
                                & \textbf{Task}                                                                                                                                                                 & \textbf{Ground truth SQL query}                                                                                                                                                                                                                                                                                                          \\
\textbf{Easy}                   
                & {List the name of teachers whose hometown \\is not “Little Lever Urban District. \\\textbf{(course\_teach)}}             & {SELECT name FROM teacher \\WHERE hometown != ``little lever urban district''}                                                                                      \\
                & {What is the abbreviation \\for airline ``JetBlue Airways'' ? \\\textbf{(flight\_2)}}                                      & {SELECT Abbreviation FROM AIRLINES \\WHERE Airline = ``JetBlue Airways''}                                                                                           \\
                & {List all the student details in \\reversed lexicographical order. \\\textbf{(student\_transcripts\_tracking)}}          & {SELECT other\_student\_details FROM Students \\ORDER BY other\_student\_details DESC}                                                                                                                                                                                                                 \\
\textbf{Medium}                                
                        & {Which airlines have less than 200 flights?\\\textbf{(flights\_2)}}                                                      & {SELECT T1.Airline FROM AIRLINES AS T1 \\JOIN FLIGHTS AS T2 ON T1.uid = T2.Airline \\GROUP BY T1.Airline HAVING COUNT(*)  200}   
                    \\
                        & {Who is the earliest graduate of the school? \\ List the first name, middle name, and last name.\\\textbf{(flights\_2)}} & {SELECT first\_name , middle\_name , last\_name \\FROM Students ORDER BY date\_left ASC LIMIT 1}                                                   \\
                        & {What are the countries having \\at least one car maker? \\ List name and id.\\\textbf{(car\_1)}}                        & {SELECT T1.CountryName , T1.CountryId \\FROM COUNTRIES \\AS T1 JOIN CAR\_MAKERS AS T2 \\ON T1.CountryId = T2.Country \\GROUP BY T1.CountryId HAVING COUNT(*) = 1} \\
\textbf{Hard}  
                                & {What are the ids and names of the \\battles that led to more than 10 \\people killed in total?\\\textbf{(battle\_death)}}                                                    & {SELECT T1.id , T1.name FROM battle AS T1 \\JOIN ship AS T2 ON T1.id = T2.lost\_in\_battle \\JOIN death AS T3 ON T2.id = T3.caused\_by\_ship\_id \\GROUP BY T1.id HAVING SUM(T3.killed)  10}                                                                                                                                             \\
                                & {What is the maximum number of times that \\a course shows up in different transcripts \\and what is that course's enrollment id?\\\textbf{(student\_transcripts\_tracking)}} & {SELECT COUNT(*) , student\_course\_id \\FROM Transcript\_Contents \\GROUP BY student\_course\_id \\ORDER BY COUNT(*) DESC LIMIT 1}                                                                                                                                                                                                      \\
                                & {What are the first names of the students who \\live in Haiti permanently or have the cell \\phone number 09700166582?\\\textbf{(student\_transcripts\_tracking)}}            & {SELECT T1.first\_name FROM students AS T1 \\JOIN addresses AS t2 \\ON T1.permanent\_address\_id = T2.address\_id \\WHERE T2.country = 'haiti' \\OR T1.cell\_mobile\_number = '09700166582’}        
                                
                                \\
{\textbf{Extra}\\\textbf{hard}}     
                                & {Which owner has paid the largest \\amount of money in total for their dogs? \\Show the owner id and zip code.\\\textbf{(dog\_kennels)}}                                      & {SELECT T1.owner\_id , T1.zip\_code FROM Owners AS T1 JOIN Dogs \\AS T2 ON T1.owner\_id = T2.owner\_id JOIN Treatments AS T3 \\ON T2.dog\_id = T3.dog\_id \\GROUP BY T1.owner\_id \\ORDER BY sum(T3.cost\_of\_treatment) DESC LIMIT 1}                                                                                                   \\
                                & {What is the area code in which the most \\voters voted?\\\textbf{(voter\_1)}}                                                                                                & {SELECT T1.area\_code FROM area\_code\_state AS T1 \\JOIN votes AS T2 ON T1.state = T2.state \\GROUP BY T1.area\_code \\ORDER BY COUNT(*) DESC LIMIT 1}                                                                                                                                                                                  \\
                                & {What is the maximum horsepower and the \\make of the car models with 3 cylinders?\\\textbf{(car\_1)}}                                                                        & {SELECT T2.horsepower , T1.Make FROM CAR\_NAMES AS T1 \\JOIN CARS\_DATA AS T2 ON T1.MakeId = T2.Id WHERE T2.cylinders = 3 \\ORDER BY T2.horsepower DESC LIMIT 1}                                                                                                                                                                         
\end{tblr}
}
\end{table*}

\edited{Table~\ref{tab:tasks3} present examples of tasks with different difficulty levels from the 48 tasks used in our study.
Table~\ref{tab:tasks3} also render the databases these tasks were operated on, as well as the ground-truth SQL queries for these tasks.}
These tasks were selected from the Spider benchmark~\cite{spider}. Spider is a large-scale, complex, and cross-domain benchmark, consisting of databases with multiple tables. It has become the de facto standard for measuring text-to-SQL models these days. Spider categorizes these tasks into four difficulty levels---easy, medium, hard, and extra hard. 
We performed a stratified random sampling on the tasks from Spider~\cite{spider}. Specifically, we selected 12 easy tasks, 12 medium tasks, 12 hard tasks, and 12 extra hard tasks, according to the difficulty classification from Spider. For each participant and each tool/variant assignment during the study, we randomly selected 2 tasks per difficulty level from the pool of 48 tasks, resulting in 8 tasks per condition. We randomized the order of the 8 tasks to counterbalance the impact of task difficulty levels (e.g., doing easy tasks first vs.~doing difficult tasks first). If a participant found a task too difficult to solve, they were allowed to skip it.

\section{User Interfaces of {\tool} and Baselines}

\label{app:ui}
This section demonstrates the user interface (UI) of baseline tools used in our user study I.

\textbf{MISP.} Given a natural language question, 
MISP may ask users multiple-choice questions to clarify which column should be considered.
If none of the listed choices are correct, users are allowed to provide their own answers.
The user's answer is used to constrain the decoding process by adjusting the probability of code tokens induced by the answer.
However, MISP directly renders the generated SQL to users without explanation. Therefore, users need to be familiar with SQL syntax to identify errors.
The official implementation of MISP on GitHub only had a command-line interface, and the original text-to-SQL model~\cite{editsql} had much lower accuracy than newer models.
To enable a fair comparison, we first created an interface for MISP, which includes everything from the {\tool} interface except the \textit{Query Explanation} view (Figure~\ref{fig:UI} \circled{D}). Then, we replaced their text-to-SQL model~\cite{editsql} with the one~\cite{smbop} used in {\tool}. Thus, the only difference between the two systems is the interaction mechanism.

\begin{figure*}[htb]
    \centering
    \includegraphics[width=\linewidth]{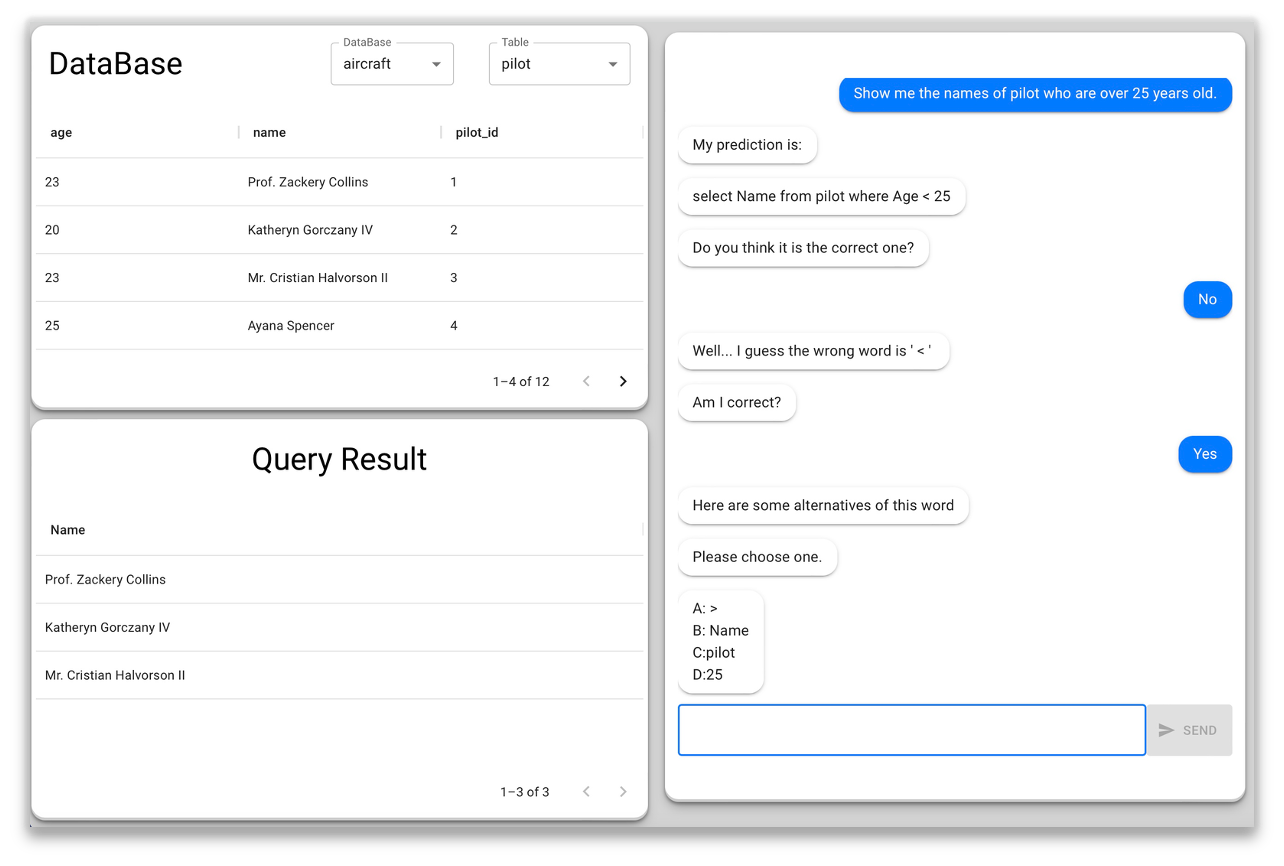}
    \caption{The UI of MISP}
    \label{fig:misp}
\end{figure*}

As shown in Figure~\ref{fig:misp}, MISP shares a similar UI as {\tool} (Figure \ref{fig:UI}). For each query task, MISP allows users to select a database, inspect data in a table, and view the query result. The main difference from {\tool} is that MISP will render a generated query in the dialog and ask users to confirm whether the generated SQL is correct or not. If the user says the generated query is not correct, MISP will proactively predict the erroneous part and ask users to select alternative generations to fix the error. However, MISP does not provide a natural language explanation of the generated SQL. Users have to read and inspect the generated SQL in the dialog on their own, which is difficult for end-users who do not understand the syntax and semantics of SQL.

\textbf{DIY.}
Given a natural language question, DIY creates a small sampled database and computes intermediate results on the samples. Furthermore, DIY maps tokens in a generated SQL query to words and phrases in the user-provided question. If the user finds an incorrect mapping (e.g., a wrong column name), they can fix it by selecting an alternative name and value from a drop-down menu.
However, users cannot give further feedback in addition to selecting alternatives at certain locations.
Since the original implementation of DIY is not publicly available, we reused the replication of DIY from Ning et al.~\cite{ning_empirical_2023} and designed a user interface similar to {\tool}. We also changed the original text-to-SQL model in Ning et al.'s implementation to the same model~\cite{smbop} of {\tool} for a fair comparison. 

\begin{figure*}[htb]
    \centering
    \includegraphics[width=\linewidth]{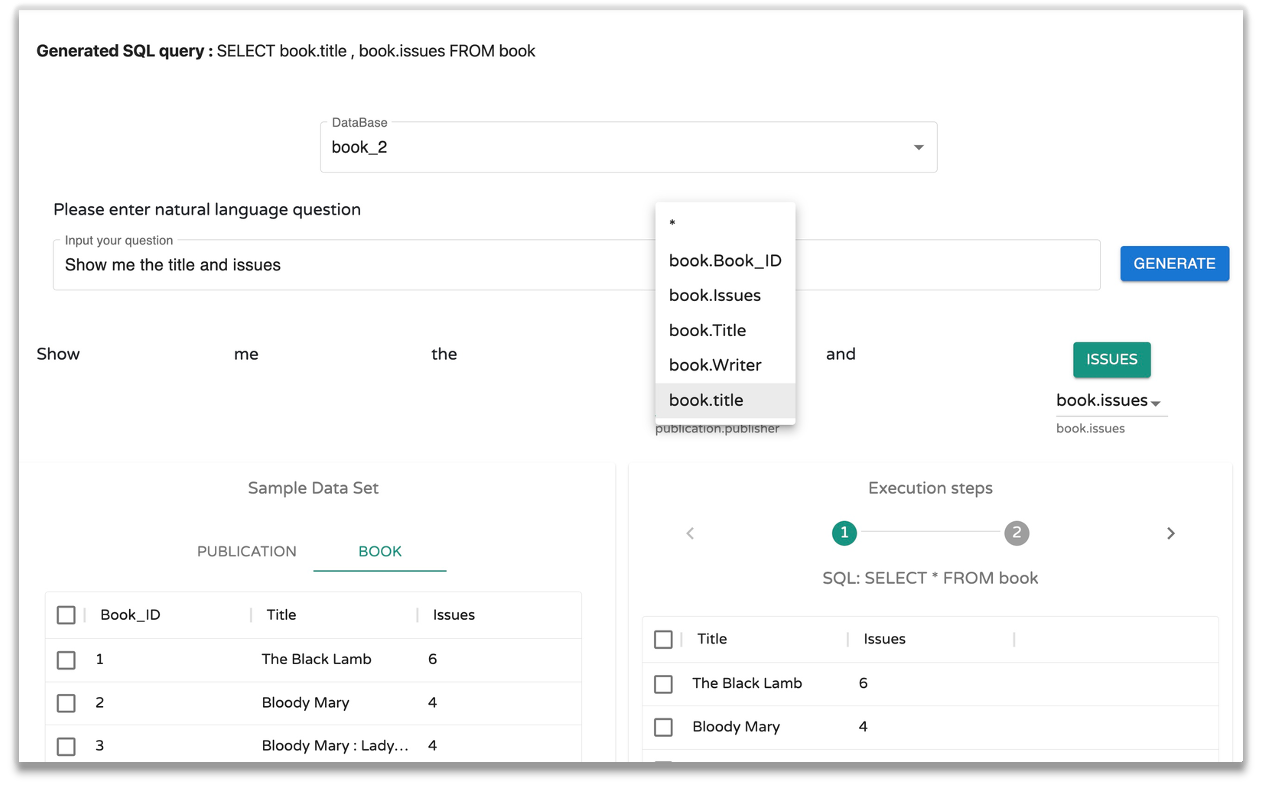}
    \caption{The UI of DIY}
    \label{fig:diy}
\end{figure*}

Figure~\ref{fig:diy} shows the UI of DIY. DIY only samples a small amount of data from a user-selected database to reduce the information overload of inspecting a large database. Users can type in a natural language question and then DIY generates a SQL query by invoking the base SQL generation model. DIY automatically matches tokens in the natural language question with tokens in the generated SQL. Each matched natural language token is augmented with a drop-down menu with alternative SQL tokens predicted by the base model. If the prediction of a token is wrong, users can click on the drop-down menu and select an alternative token to fix it. Users can examine the query result, as well as the execution steps, in the bottom right view.

\end{document}